\documentclass{aa}

\usepackage{graphicx}
\usepackage{natbib}
\usepackage{txfonts}

\begin{document}

   \title{GALACTICNUCLEUS: A high-angular-resolution $JHK_s$ imaging survey of the Galactic centre}
   \titlerunning{GALACTICNUCLEUS IV}
   \authorrunning{Nogueras-Lara et al.}

   \subtitle{IV. Extinction maps and de-reddened photometry.\thanks{The extinction maps and the catalogues described in this paper and specified in Sect.\,\ref{dupli} and Table\,\ref{final_cat} are available in electronic form at the CDS via anonymous ftp to cdsarc.u-strasbg.fr (130.79.128.5) or via http://cdsweb.u-strasbg.fr/cgi-bin/qcat?J/A+A/.}}

  \author{F. Nogueras-Lara
          \inst{1}
          \and      
          R. Sch\"odel 
          \inst{2}       
          \and
          N. Neumayer 
          \inst{1}    
          }

   \institute{
    Max-Planck Institute for Astronomy, K\"onigstuhl 17, 69117 Heidelberg, Germany
              \email{nogueras@mpia.de}
       \and 
           Instituto de Astrof\'isica de Andaluc\'ia (CSIC),
     Glorieta de la Astronom\'ia s/n, 18008 Granada, Spain                         
}
   \date{}

% \abstract{}{}{}{}{} 
% 5 {} token are mandatory
 
  \abstract
   {The extreme extinction ($A_V\sim30$\,mag) and its variation on arc-second scales towards the Galactic centre hamper the study of its stars.  Analysis of them is restricted to the near infrared (NIR) regime, where the extinction curve can be approximated by a broken power law for the $JHK_s$ bands. Therefore, it is fundamental to correct for extinction at these wavelengths in order to analyse the structure and stellar population of the central regions of our Galaxy.}
   {We aim to, (1) discuss different strategies to de-redden the photometry and check the usefulness of extinction maps to deal with variable stars; (2) build  extinction maps for the NIR bands $JHK_s$ and make them publicly available; (3) create a de-reddened catalogue of the GALACTICNUCLEUS (GNS) survey, identifying foreground stars; and (4) perform a preliminary analysis of the de-reddened $K_s$ luminosity functions (KLFs).}
   {We used photometry from the GNS survey to create extinction maps for the whole catalogue. We took red clump (RC) and red giant stars of similar brightnesses as a reference to build the maps and  we de-reddened the GNS photometry. We also discussed the limitations of the process and analysed non-linear effects of the de-reddening.  
}
{We obtained high resolution ($\sim3''$) extinction maps with low statistical and systematics uncertainties ($\lesssim5$\,\%) and computed average extinctions for each of the regions covered by the GNS. We checked that our maps effectively correct the differential extinction reducing the spread of the RC features by a factor of $\sim2$. We assessed the validity of the broken power law approach computing two equivalent extinction maps $A_H$ using either $JH$ and $HK_s$ photometry for the same reference stars and obtained compatible average extinctions within the uncertainties. Finally, we analysed de-reddened KLFs for different lines of sight and found that the regions belonging to the NSD contain a homogeneous stellar population that is significantly different from that in the innermost bulge regions.}
   
   {}


\titlerunning{GALACTICNUCLEUS. IV.}
\authorrunning{F. Nogueras-Lara et al.}

   \maketitle
%
%________________________________________________________________

 \section{Introduction}

The Galactic centre (GC) is an environment of fundamental importance for astrophysics beacuse it is the closest galactic nucleus, located at only 8 kpc from Earth \citep[e.g.][]{Gravity-Collaboration:2018aa,Do:2019aa}, and we can analyse its structure and stellar population, resolving individual stars down to milliparsec scales. It hosts a supermassive black hole at its centre, Sgr A*, which is embedded in a nuclear stellar cluster \citep[e.g.][]{Ghez:1998ad,Schodel:2002zt,Genzel:2010fk,Schodel:2014fk}. At larger scales, the nuclear stellar disc (NSD) and the central molecular zone (CMZ) outline the limits of the GC \citep[e.g.][]{Morris:1996vn,Launhardt:2002nx,Kruijssen:2014aa,Nogueras-Lara:2019ad}.

The extreme extinction towards the GC ($A_V\gtrsim30$ mag, $A_{K_s}\gtrsim2.5$ mag, e.g. \citealt{Nishiyama:2008qa,Schodel:2010fk,Nogueras-Lara:2018aa,Nogueras-Lara:2020aa}) impedes the study of its stellar populations in the optical. The extreme source crowding requires the use of high-angular-resolution observations to disentangle individual stars. Even previous studies using the VISTA Variables in the Via Lactea (VVV) survey \citep{Minniti:2010fk,Saito:2012ml}, which constitutes the most accurate up-to-date survey for the Galactic bulge, are not sufficient to properly determine the extinction variations in the innermost regions of our Galaxy \citep[e.g.][]{Gonzalez:2012aa,Gonzalez:2013aa,Surot:2020vo}. To improve this situation, the GALACTICNUCLEUS (GNS) survey \citep{Nogueras-Lara:2018aa,Nogueras-Lara:2019aa} was optimised for the requirements of the GC ($JHK_s$, high angular resolution, high dynamic range) and is currently the most complete near infrared (NIR) $JHK_s$ photometric catalogue of the GC. It provides high-angular-resolution ($\sim 0.2''$) photometry of more than 3 million stars distributed along the NSD, the inner bulge, and the transition region between them.

The extinction towards the GC varies significantly on arc-second scales \citep[][]{Scoville:2003la,Schodel:2010fk,Nogueras-Lara:2018aa} due to the presence of a significant amount of gas, dust, and molecular dark clouds associated with the CMZ \citep[e.g.][]{Dahmen:1998aa,Pierce-Price:2000aa}. In order to quantify the magnitude of the extinction variations within the field of view and to trace the presence of the dark clouds, extinction maps can be extremely useful. They inform us about the properties and distribution of the interstellar medium (ISM), given that they are directly related to the dust content. Moreover, they can be used to determine the gas column density distribution assuming a constant gas to dust ratio \citep[e.g.][]{Goodman:2009wp,Rowles:2009wt}.

To create extinction maps, it is key to understand the behaviour of the extinction curve. It has been generally assumed that the extinction curve in the NIR can be well approximated as a power law \citep[e.g.][]{Rieke:1985fq,Nishiyama:2006tx,Fritz:2011fk}. However, recent studies found a dependence of the extinction on the wavelength, which points towards a more complicated broken-power-law behaviour of the form $A_{\lambda} \propto \lambda^{-\alpha_{JH}}$ and $A_{\lambda} \propto \lambda^{-\alpha_{HK_s}}$, where $\alpha_{i}$ is the extinction index and the i-subindinces indicate their range of validity between the photometric bands $JH$ and $HK_s$ \citep{Nogueras-Lara:2018aa,Hosek:2018aa,Nogueras-Lara:2019ac,Nogueras-Lara:2020aa}. Recent work by \citet{Nogueras-Lara:2020aa}, carried out using the whole GNS survey, obtained $\alpha_{JH} = 2.44\pm0.05$ and $\alpha_{HK_s} = 2.23\pm 0.05$.

In this paper, we compare the de-reddening of NIR photometry using a star-by-star basis and an extinction map approach. Moreover, we also create the most detailed high-angular-resolution extinction maps of the GC (covering a total region of $\sim 6000$ pc$^2$) using the GNS survey. We validate the broken-power-law approach and de-redden the photometry from GNS to obtain de-reddened colour-magnitude diagrams (CMDs) of the GC. We publish the de-reddened catalogue, analyse the limitations of the de-reddening, and study $K_s$ luminosity functions (KLFs) to carry out a preliminary study of the stellar population for the different regions covered.

 \section{Data}
 
\begin{figure}
   \includegraphics[width=\linewidth]{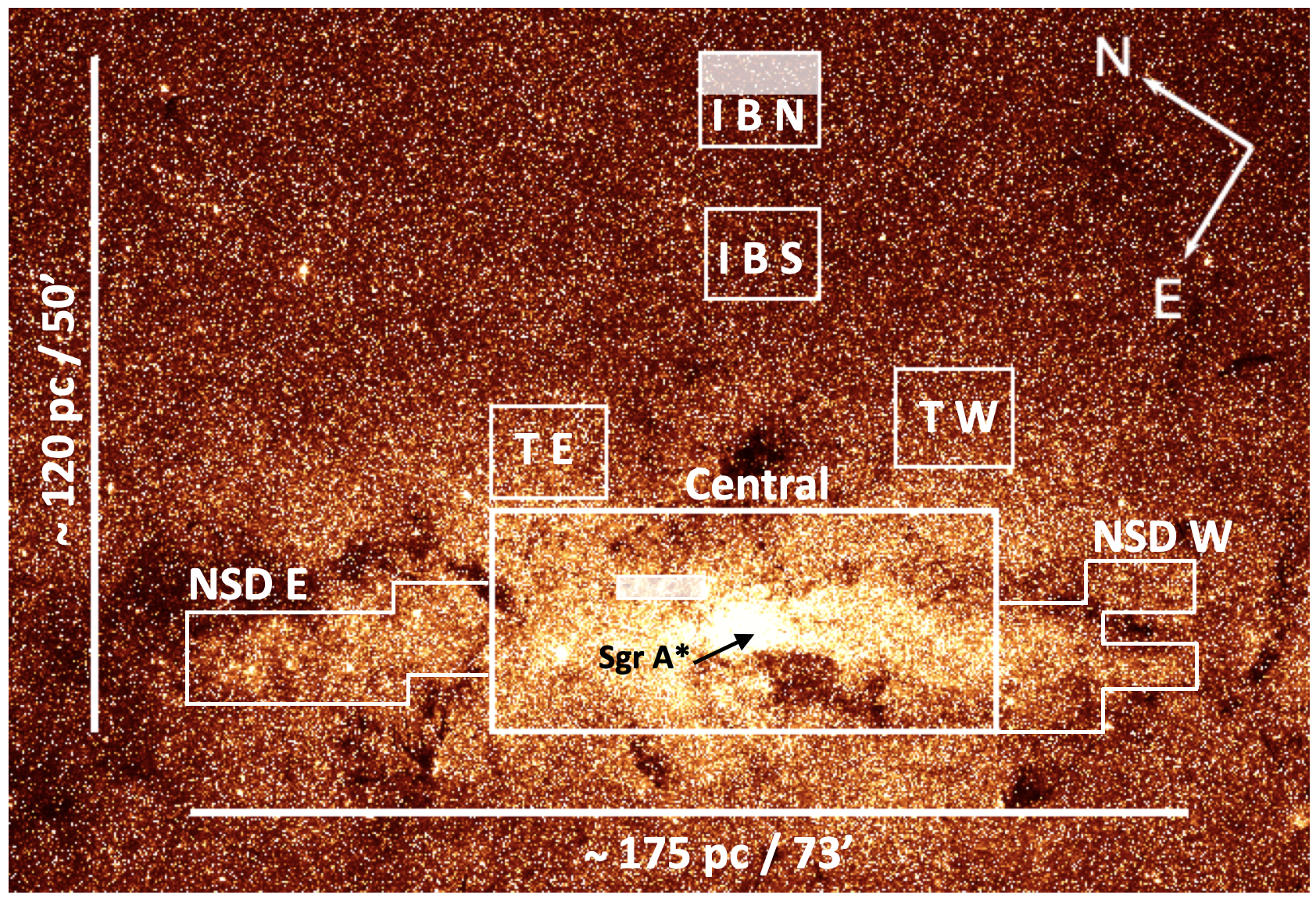}
   \caption{Scheme of the GNS survey. White labels indicate the name of each of the regions of the survey: Central, NSD East (NSD E), NSD West (NSD W), transition East (T E), transition West (T W), inner bulge South (I B S), and inner bulge North (I B N). The white rectangles in the Central and I B S regions indicate a gap in the data due to bad quality (see text for details). The background image corresponds to a 3.6 $\mu$m Spitzer/IRAC image \citep{Stolovy:2006fk}.}

   \label{GNS}
    \end{figure}

For the study presented in this paper, we used the GNS survey. This is a $JHK_s$ NIR high-angular-resolution ($\sim 0.2''$) catalogue, obtained using the HAWK-I instrument \citep{Kissler-Patig:2008uq} at the ESO Very Large Telescope unit telescope 4. It covers an area of $\sim 6000$ pc$^2$ distributed along seven regions that correspond to the nuclear stellar disc (Central, NSD East, and NSD West), two regions in the inner bulge (IB South and IB North), and two regions in the transition region between the NSD and the inner bulge (T East and T West). Figure \ref{GNS} shows a scheme of the areas covered by the survey. The GNS catalogue was calibrated using the SIRIUS/Infrared Survey Facility telescope (IRSF) GC survey \citep[e.g.][]{Nagayama:2003fk,Nishiyama:2006tx} and has a systematic zero point (ZP) uncertainty of 0.04 mag in all three bands. It includes accurate photometry of $\sim 3.3\times 10^6$ sources that reaches  5\,$\sigma$ detections for $J\sim22$, $H\sim21$, and $K_s\sim21$ mag. The photometric uncertainties are below 0.05 mag at $J\lesssim20$, $H\lesssim17$, and $K_s\lesssim16$\,mag \citep{Nogueras-Lara:2019aa}.

\subsection{Fixing duplicated stars}
\label{dupli}

We noticed that some stars are duplicated in the final GNS catalogue. Slightly different positions for different detectors, pointings and/or bands, or the presence of very close neighbouring stars might explain this problem. The observing strategy to obtain the GNS survey includes 49 pointings. Each pointing is composed of the combination of the four HAWK-I detectors, which are separated by a cross-shaped gap between them. To cover this gap, we applied random jittering to the observations and computed the photometry on an individual detector basis \citep[for further details, see][]{Nogueras-Lara:2019aa}. In this way, there are many overlapping regions between different detectors and pointings. The catalogues for each band were obtained combining all the detections from each independent detector and then from different pointings. We lastly combined all three bands into a final catalogue. Therefore, we believe that some stars escaped to be properly combined throughout the process.

We identified the duplicated stars and combined them into single sources. We checked whether they were detected in one or several bands and combined the coordinates and photometry accordingly, following the catalogue scheme shown in Table 2 of \citet{Nogueras-Lara:2019aa}. The uncertainties were quadratically propagated from the individual detections. To identify these stars in the corrected catalogue, we modified the parameters $i_J$, $i_H$, and $i_{K_s}$, which correspond to the multiple detections in overlapping pointings for each band \citep[see Sect. 4.4 for further details; ][]{Nogueras-Lara:2019aa}, and set them to $-1$.

The number of duplicated stars is very low, so they do not affect any of the previous studies carried out using the GNS survey. Namely, for each of the catalogue regions, we found a rate of duplicated stars of: Central $\sim0.002\,\%$,  NSD East $\sim0.01\,\%$, NSD West $\sim0.02\,\%$, T East $\sim0.04\,\%$, T West $\sim0.01\,\%$, IB South $\sim0.02\,\%$, and IB North $\sim0.04\,\%$.

\subsection{Colour-magnitude diagrams}
\label{contamination}

\begin{figure*}
   \includegraphics[width=\linewidth]{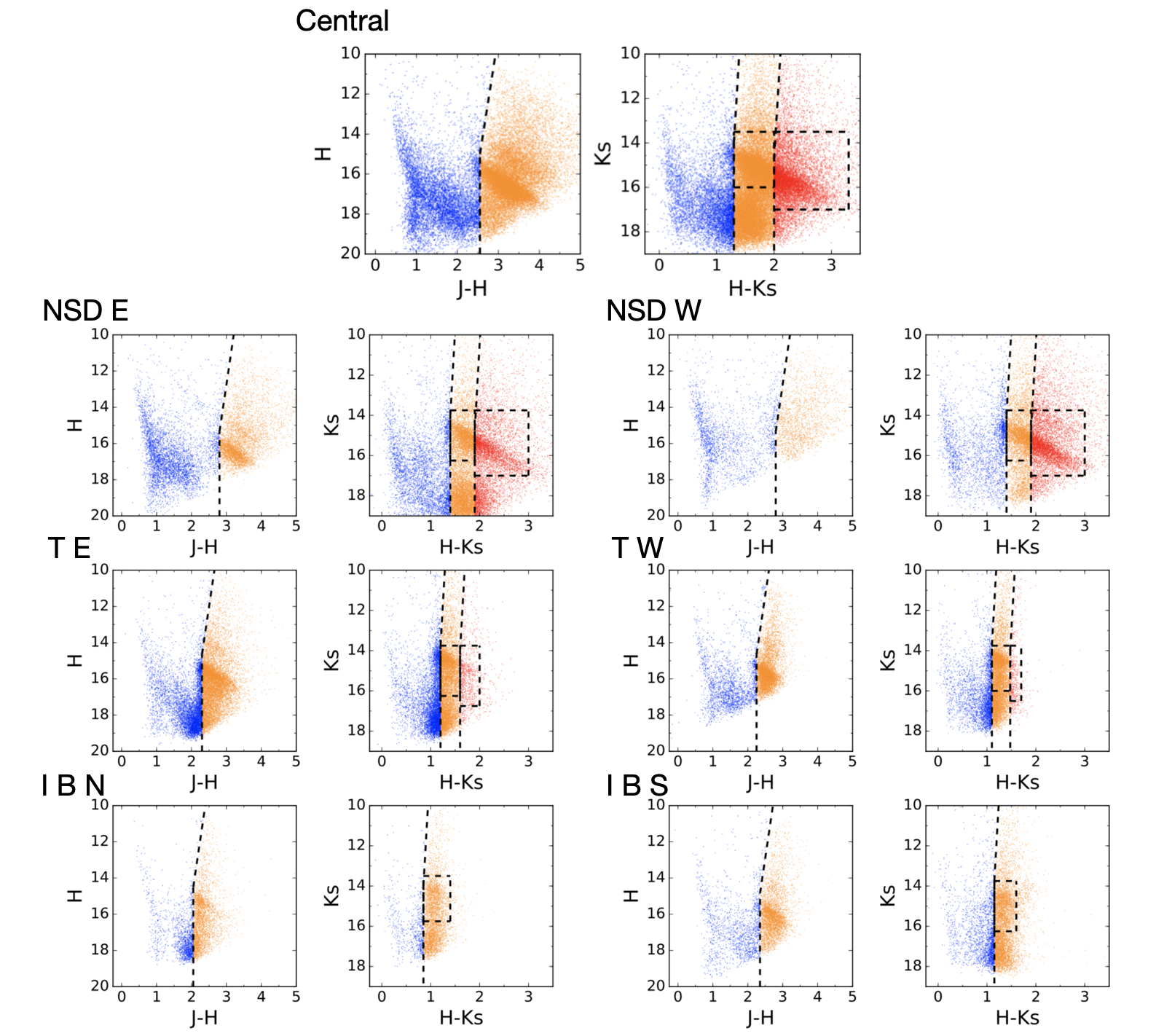}
   \caption{Colour-magnitude diagrams for each region of the GNS survey. The dashed lines represent the cuts applied in this work between the foreground population (in blue) and stars that we consider to lie inside the GC, behind two different layers of extinction (orange and red). The extinction layers were selected following the methodology explained in Sect.\,\ref{sel_criterion}. The dashed rectangles indicate the selection of reference stars (mainly RC stars) to create the extinction maps.  Only two layers are defined for the IBN and IBS regions. The low number of stars present for the NSD W region is due to the lower than average quality of the data for that region. Given the high number of stars, only a fraction of them is represented to not overcrowd the plots.}

   \label{fig_star_sel}
    \end{figure*}

Figure \ref{fig_star_sel} shows the colour-magnitude diagrams (CMDs) $H$ versus $J-H$ and $K_s$ versus $H-K_s$ for the different regions of the GNS survey. The CMDs are largely complete in the red clump (RC) region, which is key to produce extinction maps \citep[e.g.][]{Nogueras-Lara:2018aa,Nogueras-Lara:2020aa}. Red clump stars are red giant stars in their helium burning phase whose intrinsic properties are well known \citep[e.g.][]{Girardi:2016fk}. The over-densities around $H\sim 16-17$\,mag and $K_s\sim15-16$\,mag correspond to the RC features.

To produce extinction maps, we removed foreground stars using the significantly different extinction between the stars from the Galactic disc and the GC \citep[e.g.][]{Nogueras-Lara:2019aa}. We applied a colour cut, $J-H\sim2.35$ or $H-K_s\sim1.1$\,mag, depending on the CMD used (see Sect.\,2 in \citealt{Nogueras-Lara:2021uz}). 

We also defined another colour cut to remove stars belonging to the Galactic bulge in the GNS regions covering the NSD (Central, NSD E, NSD W, T E, and T W). This is possible because the stellar population from the GC lies behind a dense screen of extinction that may be related to the CMZ \citep[e.g.][]{Schultheis:2009tg}. We adopted $H-K_s\sim1.3$\,mag following previous results on the inner bulge \citep[e.g.][]{Nogueras-Lara:2018ab, Nogueras-Lara:2019ad,Schultheis:2021wf}. Nevertheless, we treated each region in a different way due to the variation of the absolute extinction between regions. Given that we produced independent extinction maps using $JH$ and $HK_s$ bands, we excluded the foreground population considering independent cuts in both CMDs $H$ versus $J-H$ and $K_s$ versus $H-K_s$. Figure\,\ref{fig_star_sel} shows the cuts applied to remove the foreground population in each case. We estimate that the possible contamination from stars belonging to the Galactic bulge/bar is below 25\% using the model created by  \citet{Schultheis:2021wf}, which applied similar colour cuts to analyse the NSD. In particular, they modelled the NSD, the bulge/bar, and the nuclear star cluster (NSC) using the results from \citet{Launhardt:2002nx}, \citet{Chatzopoulos:2015yu}, and \citet{Sormani:2020aa}.

We adopted a two-layers approach to create the extinction maps to overcome the extreme differential extinction in the GC (see Sect.\,\ref{sel_criterion}). Figure\,\ref{fig_star_sel} indicates the criteria to assign the stars to each of the presumed extinction layers.

\section{Extinction maps}

We created extinction maps for all the regions in the GNS catalogue following the approach described in detail in Appendix\,\ref{ext_map_method}. We obtained extinction maps $A_{J,\ JH}$ and $A_{H,\ JH}$, and $A_{H,\ HK_s}$ and $A_{K_s,\ HK_s}$, using $JH$ and $HK_s$ photometry, as indicated by the subindices. We assumed the extinction curve $A_J/A_H=1.87\pm0.03$ and $A_H/A_{K_s} =1.84\pm0.03$ derived in \citet{Nogueras-Lara:2020aa}. We computed associated uncertainty maps using a Jackknife algorithm, estimating the extinction variation for a given pixel and  systematically leaving out each of the stars used for its extinction calculation \citep[for further details see Sect.\,7 in][]{Nogueras-Lara:2018aa}. The systematic uncertainties were estimated by analysing the fluctuations of the extinction maps when varying the quantities involved in their calculation within their uncertainties. These are: $A_J/A_H=1.87\pm0.03$ and $A_H/A_{K_s} =1.84\pm0.03$ (depending on the computed map), the ZP photometric uncertainty of the GNS data \citep[0.04\,mag for $JHK_s$)][]{Nogueras-Lara:2019aa}, and the intrinsic colours $JH$ and $HK_s$ for RC stars and their associated uncertainties.

\subsection{Intrinsic colour of red clump stars}
\label{intrinsic}

The reference stars used to compute the extinction maps are mainly RC stars with a fraction of other red giant stars (red giant branch bump, RGBB, stars among others, e.g. \citealt{Nogueras-Lara:2018ab}), whose intrinsic colours are very similar. Therefore, we computed their intrinsic colours $(J-H)_0$ and $(H-K_s)_0$ using RC stars as a reference. We estimated these intrinsic colours and their uncertainties with three independent methods:

(1) We selected stars in the RC features from the synthetic model created in Sect.\,\ref{synt} for the NSD and estimated average intrinsic colours $(J-H)_0=0.57\pm0.04$\,mag and $(H-K_s)_0=0.11\pm0.01$, where the uncertainties correspond to the standard deviation of the measurements.\\

(2) We used the values $(J-H)_0=0.47$\,mag and $(H-K_s)_0=0.09$\,mag computed following the method described in Sect.\,3.2 of \citet{Nogueras-Lara:2020aa}, using a Kurucz stellar model \citep{Kurucz:1993fk} for a RC star (effective temperature of 4750\,K, twice solar metallicity, and $log g = +2.5$) and the corresponding HAWK-I filters.\\

(3) Finally, we used the recent observational results $(J-H)_0=0.52\pm0.04$ mag and $(H-K_s)_0=0.10\pm0.03$\,mag from Table 3 in \citet{Plevne:2020aa}, obtained for metal rich RC stars. The metallicity of the used metal rich RC stars is slightly slower than that expected for the GC \citep[e.g.][]{Feldmeier-Krause:2017kq,Schultheis:2019aa,Nogueras-Lara:2019ad}, but is appropriate to obtain an estimation of the intrinsic colours by comparing the values with the previous ones.\\

We combined the previous values and estimated the uncertainties using the standard deviation of the distributions. The results are $(J-H)_0=0.52\pm0.04$\,mag and $(H-K_s)_0=0.10\pm0.01$\,mag.

\subsection{Extinction layers}
\label{sel_criterion}

Building an extinction map involves the simplification of an actual tridimensional structure into a 2D map that corrects for extinction. In the GC, where the differential extinction is extreme and significantly varies along the line of sight \citep[e.g.][]{Nishiyama:2006tx,Nishiyama:2009oj,Nogueras-Lara:2018aa}, this is even more complex. Moreover, the steep rise of extinction towards short wavelengths in the NIR (>5\,mag higher in $J$ than in $K_s$) implies that the number of stars detected is significantly lower in $J$ than in $H$ and $K_s$ \citep[$\sim20\%$ of the stars were detected in $J$; $\sim65\%$, in $H$; and $\sim90\%$, in $K_s$, ][]{Nogueras-Lara:2019aa}. Therefore, the catalogue reaches stars in $H$ and $K_s$ that are too extinguished to be detected in $J$. 

In order to create extinction maps to de-redden the photometry in a consistent way, we decided to take a two-layers approach that allowed us to account for the variation of the extinction along a given line of sight and also to avoid over-corrections in the case of the $J$ band. The two layers defined do not necessarily correspond to real extinction screens but are more related to the incompleteness of the $J$ band and the robustness of the de-reddening of the stellar population. In this way, and given that the stars used to create the extinction maps mainly correspond to the NSD (see Sect.\ref{contamination}), both extinction layers are expected to lie at the same distance. The difference in extinction and obscuring material is then due to the extreme differential extinction and the 'granularity' characterising the NSD and the CMZ \citep[e.g.][]{Gosling:2006eq}.

The first layer of extinction is defined considering the stars detected in all three bands ($JHK_s$), whereas the second one is based on the stars that are not detected in the $J$ band. The colour cut to distinguish the first and the second extinction layers in the CMD $K_s$ versus $H-K_s$ was chosen using the $HK_s$ counterparts of stars detected in the $J$ band. In this way, we plotted those stars in the CMD $K_s$ versus $H-K_s$ and defined the cut for the second layer (see Fig.\,\ref{fig_star_sel}). Given the lower completeness of the $J$ band and the intrinsic scatter of the data, not all the stars belonging to the first layer in the CMD $K_s$ versus $H-K_s$ exactly correspond to the stars in the same layer in CMD $H$ versus $J-H$. Because of the steep extinction curve, completeness is a strong function of wavelength. This results in more stars belonging to the first layer detected towards faint magnitudes for the CMD $K_s$ versus $H-K_s$. For the Central region in particular, around two times more stars are included in the first layer for the CMD $K_s$ versus $H-K_s$ than for the $H$ versus $J-H$ one.

For the first layer we computed four extinction maps. Using the ratio $A_J/A_H$ and the $JH$ photometry, we created a $J$- and an $H$-map. Using the $A_H/A_{K_s}$ and the $HK_s$ photometry, we created an $H$- and a $K_s$-map. For the second layer we computed two extinction maps, corresponding to the $H$ and $K_s$ bands, using the extinction ratio $A_H/A_{K_s}$.

The reference stars to create each of the layers were selected in the CMDs $K_s$ versus $H-K_s$ (for each region), given that the vast majority of stars within the two layers are present there. For the first extinction layer, we considered all the stars observed in the $J$ band that have a counterpart in $H$ and $K_s$. To select the limit between both layers, we considered the cut where the stars are observed in all three bands or only in $H$ and $K_s$. Figure \ref{fig_star_sel} shows each of the established divisions that are depicted by the black dashed rectangles. The lower $K_s$ cut of the selected stars for the second layer is fainter given that the extinction is higher here, producing redder stars with fainter $K_s$ magnitudes.

The situation is simpler for the regions belonging to the inner bulge (I B S and I B N), and a single-layer approach is sufficient. This is due to the lower crowding and extinction, together with the less significant differential extinction \citep[e.g.][]{Nogueras-Lara:2018ab,Nogueras-Lara:2020aa}.

\subsection{Results}

\begin{figure*}[t!]
   \includegraphics[width=\linewidth]{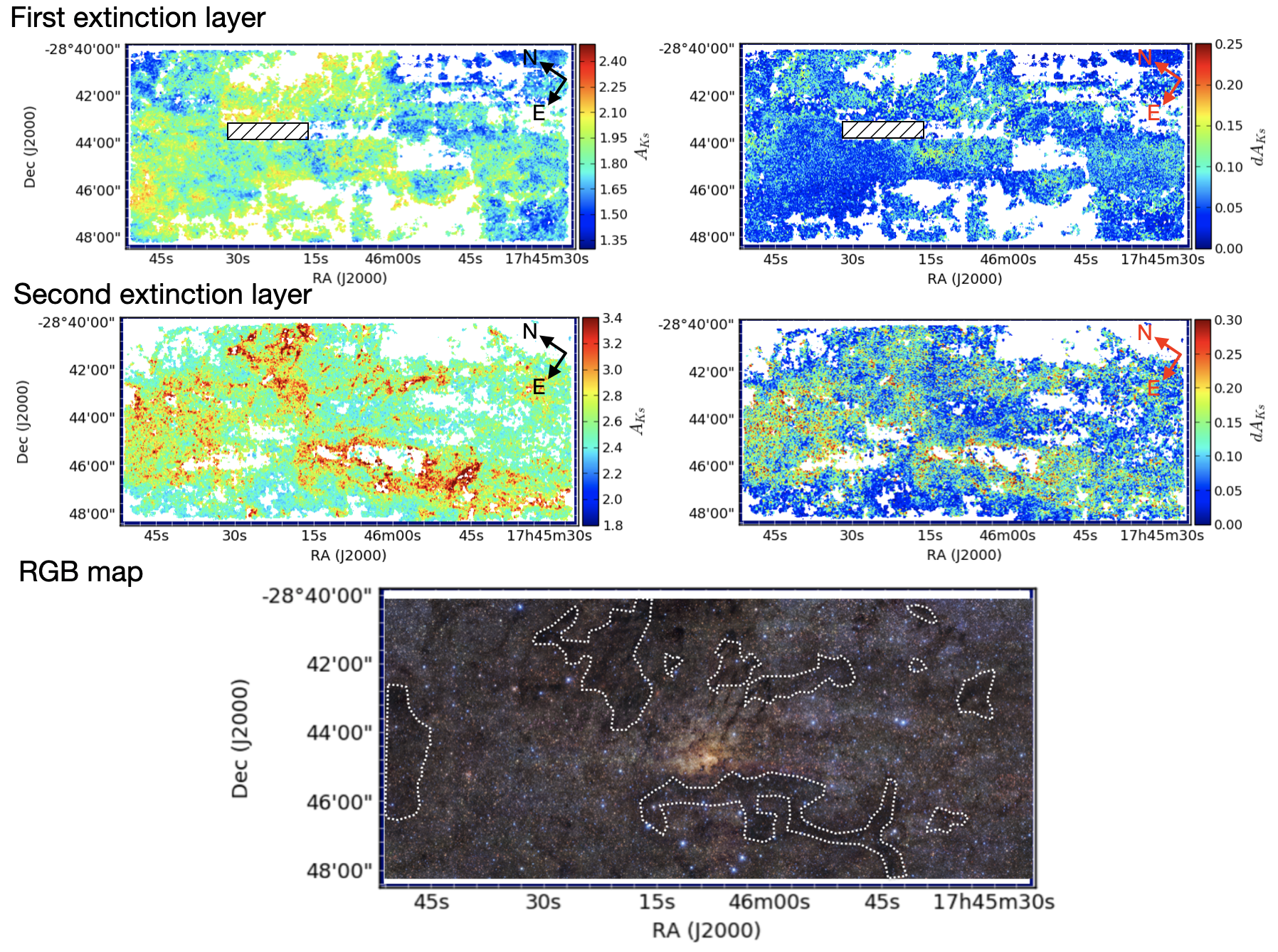}
   \caption{Extinction maps $A_{Ks}$ (first column) and associated uncertainties (second column) for the Central region of the GALACTINUCLEUS survey. The scales are different for each panel according to the values of the maps and they are indicated by the colour bar. The dashed rectangles on the upper panels indicate a region where no $J$-band data were available due to bad data quality \citep[for further details, see][]{Nogueras-Lara:2019aa}. White regions indicate that there were not enough stars to assign a value. The lower panel is a false colour image using $JHK_s$ bands. The dashed white lines indicate dark clouds that can be identified in the extinction map corresponding to the second layer.}

   \label{map_extinction_central}
    \end{figure*}

Figure \ref{map_extinction_central} shows the $A_{K_s}$ maps and their associated uncertainties obtained for the Central region of the GNS survey, as an example. The last panel of the figure shows the false colour ($JHK_s$) image of the analysed region, where it is possible to establish some correlation between the extinction maps and the dark regions in the image where the stellar density is lower. This correlation is much stronger for the second reddening layer, given that the first one correspond to stars whose extinction is lower. On the other hand, there is also some correlation between both extinction layers as it is not possible to completely distinguish stellar populations without any overlap \citep[e.g.][]{Nogueras-Lara:2018aa}. The error maps show relatively constant uncertainties that are somewhat larger in regions where there are stars that might have different extinctions and are close in projection.

We observed the strong variations caused by the differential extinction in both layers; they are  more significant and vary on smaller scales for the second extinction layer. This is in agreement with the variations found for the central pointing of the GNS survey by \citet{Nogueras-Lara:2018aa} and corresponds to the high level of complex structure or 'granularity' observed by \citet{Gosling:2006eq}, associated to extinction variations with a characteristic scale of $5-15''$.

All the maps obtained for the GNS regions are made publicly available at the Centre de Données astronomiques de Strasbourg (CDS). Table \ref{prop_cen} indicates the main properties of the computed maps. We computed the uncertainties for each of the maps and ended up with statistical and systematic uncertainties $\lesssim5\,\%$ and $\lesssim6\,\%$, respectively. We also computed the average extinction for the stars in each layer and its variation for a given band via its standard deviation (Table\,\ref{prop_cen}).

\begin{table*}[t!]
\caption{Properties of the extinction maps obtained for the Central region.}
\label{prop_cen} 
\begin{center}
\def\arraystretch{1.4}
\setlength{\tabcolsep}{3.8pt}

\begin{tabular}{cccccc|cccccc}
 &  &  &  &  & \multicolumn{1}{c}{} &  &  &  &  &  & \tabularnewline
\cline{4-9} \cline{5-9} \cline{6-9} \cline{7-9} \cline{8-9} \cline{9-9} 
 &  &  & Regions  & Band  & \multicolumn{1}{c}{$\bar{A}_{\lambda}$ } & $\sigma\bar{A}_{\lambda}$  & $\Delta_{stat}\bar{A}_{\lambda}$  & $\Delta_{syst}\bar{A}_{\lambda}$ &  &  & \tabularnewline
 &  &  &  &  & \multicolumn{1}{c}{(mag) } & (mag)  & (mag)  & (mag)  &  &  & \tabularnewline
\cline{4-9} \cline{5-9} \cline{6-9} \cline{7-9} \cline{8-9} \cline{9-9} 
 &  &  &  & $J^{1}$  & \multicolumn{1}{c}{5.80 } & 0.49  & 0.21  & 0.19  &  &  & \tabularnewline
 &  &  &  & $H^{1}$  & \multicolumn{1}{c}{3.10 } & 0.26  & 0.11  & 0.13  &  &  & \tabularnewline
 &  &  & Central & $H^{1}$  & \multicolumn{1}{c}{3.35} & 0.26  & 0.12  & 0.14  &  &  & \tabularnewline
 &  &  &  & $K_{s}^{1}$  & \multicolumn{1}{c}{1.82 } & 0.14  & 0.06  & 0.10  &  &  & \tabularnewline
 &  &  &  & $H^{2}$  & \multicolumn{1}{c}{4.89 } & 0.44  & 0.19  & 0.15  &  &  & \tabularnewline
 &  &  &  & $K_{s}^{2}$  & \multicolumn{1}{c}{2.66 } & 0.24  & 0.10 & 0.11  &  &  & \tabularnewline
\hline 
\hline 
Regions  & Band  & $\bar{A}_{\lambda}$  & $\sigma\bar{A}_{\lambda}$  & $\Delta_{stat}\bar{A}_{\lambda}$  & $\Delta_{syst}\bar{A}_{\lambda}$ & Regions  & Layer  & $\bar{A}_{\lambda}$  & $\sigma\bar{A}_{\lambda}$  & $\Delta_{stat}\bar{A}_{\lambda}$  & $\Delta_{syst}\bar{A}_{\lambda}$ \tabularnewline
 &  & (mag)  & (mag)  & (mag)  & (mag)  &  &  & (mag)  & (mag)  & (mag)  & (mag) \tabularnewline
\hline 
 & $J^{1}$  & 5.83  & 0.40  & 0.18  & 0.18  &  & $J^{1}$  & 5.53  & 0.44  & 0.20  & 0.18\tabularnewline
 & $H^{1}$  & 3.12  & 0.21  & 0.10  & 0.13  &  & $H^{1}$  & 2.96  & 0.24  & 0.11  & 0.13\tabularnewline
NSD E  & $H^{1}$  & 3.40  & 0.18  & 0.10  & 0.14  & NSD W  & $H^{1}$  & 3.46  & 0.18  & 0.10  & 0.14\tabularnewline
 & $K_{s}^{1}$  & 1.85  & 0.10  & 0.05  & 0.10  &  & $K_{s}^{1}$  & 1.88  & 0.10  & 0.05  & 0.09\tabularnewline
 & $H^{2}$  & 4.82  & 0.37  & 0.22  & 0.15  &  & $H^{2}$  & 4.74  & 0.39 & 0.20  & 0.15\tabularnewline
 & $K_{s}^{2}$  & 2.62  & 0.20  & 0.12  & 0.11  &  & $K_{s}^{2}$  & 2.58  & 0.21  & 0.11  & 0.11\tabularnewline
\hline 
 & $J^{1}$  & 4.63  & 0.39  & 0.14  & 0.17  &  & J$^{1}$  & 4.36  & 0.25  & 0.10  & 0.17\tabularnewline
 & $H^{1}$  & 2.47  & 0.21  & 0.08  & 0.12  &  & $H^{1}$  & 2.33  & 0.13  & 0.06  & 0.11\tabularnewline
T E  & $H^{1}$  & 2.76  & 0.16  & 0.08  & 0.14  & T W  & $H^{1}$  & 2.51  & 0.13  & 0.06  & 0.13\tabularnewline
 & $K_{s}^{1}$  & 1.50  & 0.09  & 0.04  & 0.09  &  & $K_{s}^{1}$  & 1.36  & 0.07  & 0.03  & 0.08\tabularnewline
 & $H^{2}$  & 3.59  & 0.12  & 0.10  & 0.14 &  & $H^{2}$  & 3.18  & 0.06  & 0.05  & 0.14\tabularnewline
 & $K_{s}^{2}$  & 1.95  & 0.06  & 0.05  & 0.10  &  & $K_{s}^{2}$  & 1.73  & 0.03  & 0.03  & 0.09\tabularnewline
\hline 
 & $J^{1}$  & 3.89  & 0.40  & 0.11  & 0.16  &  & $J^{1}$  & 4.67  & 0.35  &0.11  & 0.17\tabularnewline
I B N  & $H^{1}$  & 2.08  & 0.21  & 0.06  & 0.10  & I B S  & $H^{1}$  & 2.50  & 0.19  & 0.06  & 0.12\tabularnewline
 & $H^{1}$  & 2.17  & 0.22  & 0.06  & 0.13  &  & $H^{1}$  & 2.74  & 0.19  & 0.06  & 0.14\tabularnewline
 & $K_{s}^{1}$  & 1.18  & 0.12  & 0.03  & 0.08  &  & $K_{s}^{1}$  & 1.49  & 0.11  & 0.04  & 0.09\tabularnewline
\hline 
\end{tabular}

\end{center}
\footnotesize
\textbf{Notes.} The super-indices 1 and 2 indicate the first and the second extinction layer, respectively. The mean extinction for the stars corresponding to each extinction map is indicated by $\bar{A}_\lambda$. The associated standard deviation, $\sigma \bar{A}_\lambda$, shows the variation of the extinction within a layer for a given band. The symbols $\Delta_{stat} \bar{A}_\lambda$ and $\Delta_{syst} \bar{A}_\lambda$ indicate the average statistical uncertainty computed using a Jackknife algorithm for a given band and layer, and the systematic uncertainty of the corresponding extinction map, respectively. In all the cases the first $H^1$ corresponds to the extinction layer obtained using the $JH$ bands, whereas the second one was computed using $HK_s$. The results obtained for NSD W are less reliable due to bad data quality caused by bad weather conditions.

 \end{table*}

The maps for the Central region and the IBN include significant gaps (see Fig. \ref{map_extinction_central} for the Central region), given that two of the pointings that form part of these regions were observed under very bad weather conditions and were therefore not included in the GALACTINUCLEUS catalogue.

\section{Quality of the extinction maps and extinction curve assessment}

%\subsection{Extinction curve assessment}

\begin{figure}[t!]
   \includegraphics[width=\linewidth]{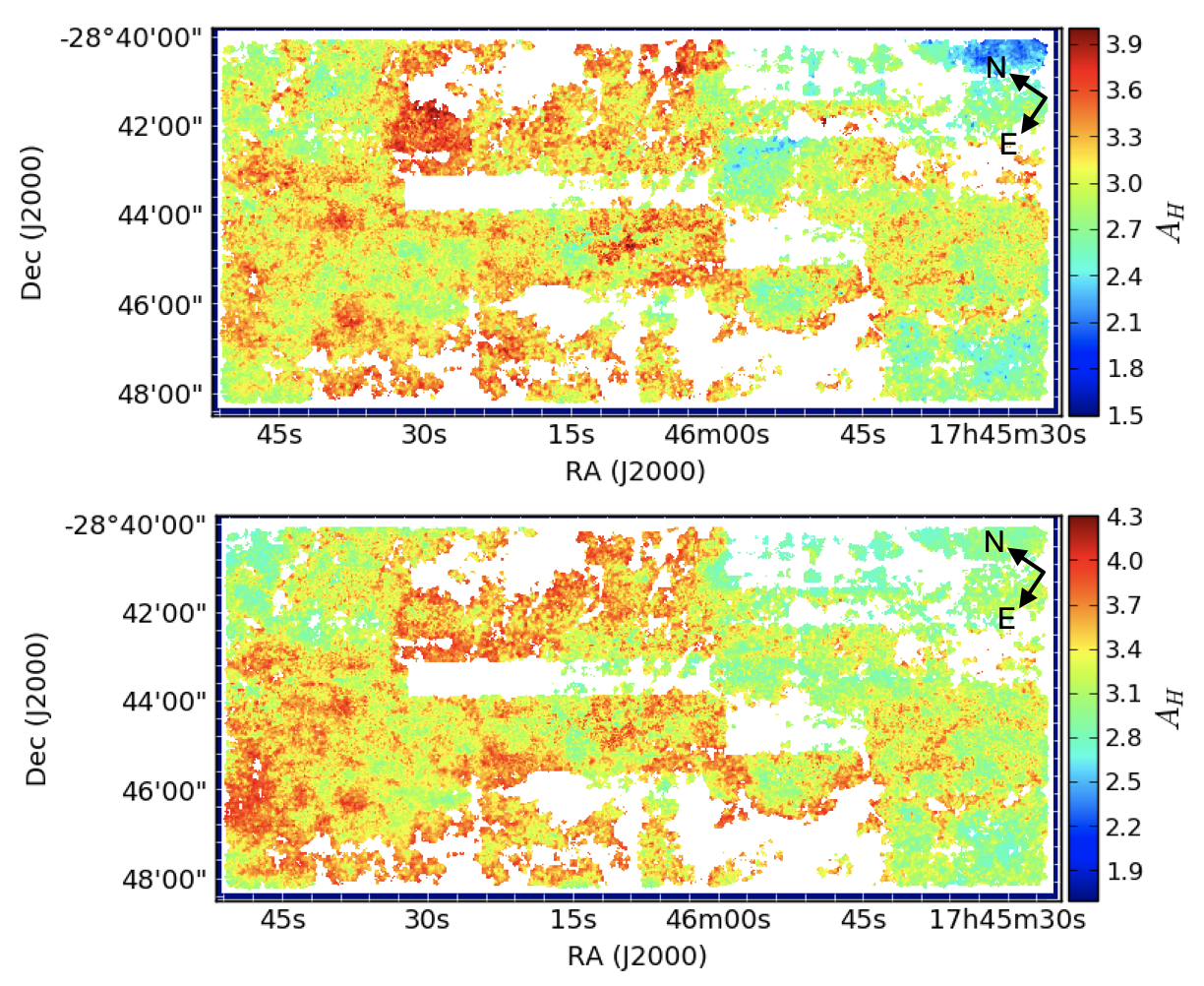}
   \caption{Extinction maps, $A_{H}$, corresponding to the first extinction layer of the Central region, computed using $JH$ (upper panel) and $HK_s$ bands (lower panel).}

   \label{AH_maps}
    \end{figure}

Since we created two independent extinctions maps $A_H$ for the first extinction layer, we could compare them with each other to assess their reliability as well as the consistency of the extinction curve that we had assumed. Figure\,\ref{AH_maps} shows both maps for the Central region.  Qualitatively, they share the main extinction features without important disagreement within the uncertainties ($dA_H\sim 0.20$\,mag, estimated by propagating the statistical and systematics uncertainties in Table\,\ref{prop_cen}). Moreover, we computed the average extinction obtained for the stars corresponding to each map (Table\,\ref{diff_ext_layers}) and concluded that the values agree within the uncertainties. The only exception is the NSD W region where the lower than average quality of the data for this field \citep[see Table A.3. in][]{Nogueras-Lara:2019aa} makes the results unreliable. In general, we observed some tendency towards larger values for the mean extinctions computed by using the $HK_s$ bands (so we used slightly different scales for each map in Fig. \ref{diff_ext_layers}). We believe that a possible variation of the ZP, within the systematic uncertainties \citep[$\sim0.04$\,mag in all three bands,][]{Nogueras-Lara:2019aa}, might explain this systematic variation. In particular, considering the mean $A_H$ values obtained for the Central region (Table\,\ref{diff_ext_layers}) and assuming that the ZP is shifted +0.04\,mag for $J$ and $K_s$ and -0.04\,mag for $H$, would reduce the obtained difference of $A_{H,\ HK_s}-A_{H,\ JH}=0.25$\,mag to $A_{H,\ HK_s}-A_{H,\ JH}\sim0.02$\,mag.

\begin{table}[t!]
\caption{Average extinction $A_H$ for the first extinction layer $A_H^1$.}
\label{diff_ext_layers} 
\begin{center}
\def\arraystretch{1.4}
\setlength{\tabcolsep}{3.8pt}
\begin{tabular}{ccc}
 &  & \tabularnewline
\hline 
\hline 
Regions & $A_{H,\ JH}$ & $A_{H,\ HK_s}$\tabularnewline
 & (mag) & (mag) \tabularnewline
\hline 
Central & 3.10$\pm$0.17 & 3.35$\pm$0.18\tabularnewline
NSD E & 3.12$\pm$0.16 & 3.40$\pm$0.17\tabularnewline
NSD W & 2.96$\pm$0.17 & 3.46$\pm$0.17\tabularnewline
T E & 2.47$\pm$0.14 & 2.76$\pm$0.16\tabularnewline
T W & 2.33$\pm$0.14 & 2.51$\pm$0.14\tabularnewline
I B N & 2.08$\pm$0.12 & 2.17$\pm$0.14\tabularnewline
I B S & 2.50$\pm$0.13 & 2.74$\pm$0.15\tabularnewline
\hline 
\end{tabular}

\end{center}
\footnotesize
\textbf{Notes.} The symbols $A_{H,\ JH}$ and $A_{H,\ HK_s}$ indicate whether the $A_H$ values correspond to the extinction map created using the photometric bands $JH$ or $HK_s$. 

 \end{table}

%We ended up with $A_{H\_ {JH}} = 3.05\pm0.18$ mag and $A_{K_s\_{HK_s}} = 3.40\pm0.17$, that are compatible within the uncertainties. 

\section{Comparison with previous work}    

\subsection{Previous 2D extinction maps}

Previous extinction maps for the GC region were limited to the photometric surveys covering this region. In this way, they are normally affected by low photometric completeness that impedes a good coverage of the RC feature. In particular, \citet{Schultheis:1999te,Schultheis:2009tg} used the Deep Near-Infrared Survey of the southern sky (DENIS) survey and Spitzer data to obtain extinction maps with resolutions $\sim2'$. In spite of the very different resolutions ($\sim3''$ for the extinction maps computed in this paper), we compared the extinction map obtained in \citet{Schultheis:2009tg} with our results. We used all the stars in the Central region of the GNS survey and obtained their associated extinction using the combined extinction map from \citet{Schultheis:2009tg} (see their Sect. 3.2 for further details). Averaging over the obtained values, we ended up with $A_{K_s}=2.74\pm0.72$\,mag, where the uncertainty corresponds to the standard deviation of the distribution. We used the same stars and also computed the associated extinction using our two-layers approach. We obtained a mean value of $A_{K_s}=2.30\pm0.46$\,mag, where the uncertainty was estimated via the standard deviation of the measurements. The somewhat larger value obtained when using the extinction map from \citet{Schultheis:2009tg} can be explained due to the different extinction curve used to compute the extinction \citep{Glass:1999vp}. Converting the extinction value using the extinction curve by \citet{Nogueras-Lara:2020aa} gives a value of $A_{K_s}=2.29\pm0.60$\,mag, which perfectly agrees with our result within the uncertainties.

More recently, the use of the VVV survey \citep{Minniti:2010fk,Saito:2012ml} improved the situation and allowed us to use RC stars to obtain high resolution extinction maps ($\sim2'$) for the whole Galactic bulge. Nevertheless, the high crowding for the GC region limits the detection of stars with the VVV survey \citep[see Fig.\,12 in ][]{Nogueras-Lara:2019aa}, so the associated extinction map does not prove adequate to de-redden the GC stellar population. In spite of it, we attempted to use the extinction map computed by \citet{Gonzalez:2012aa} using the VVV survey. We used the BEAM\footnote{http://mill.astro.puc.cl/BEAM/calculator.php} (Bulge Extinction And Metallicity) calculator and a selection of $\sim1$\,\% of the stars in the Central region of the GNS survey ($\sim20,000$ stars). We could not use more stars due to  the limit on the number of stars to compute the extinction using the BEAM calculator. Given the limitations of the VVV survey for the very central region of the Galaxy, the extinction was computed for only  $\sim 60$\,\% of the total number of stars. We used the extinction curves by \citet{Nishiyama:2009ph} and \citet{Cardelli:1989kx} and obtained $A_{K_s}=2.18\pm0.24$ and $A_{K_s}=2.85\pm0.31$\,mag, respectively. The uncertainties were estimated as the standard deviation of the distributions. The significantly different values obtained illustrate how different extinction curves influence the determination of the absolute extinction \citep[e.g.][]{Nogueras-Lara:2019ac}. To better compare the results with the work presented here, we translated these values into $A_{K_s}=1.69\pm0.18$\,mag using the extinction ratio $A_J/A_{K_s}=3.44$ obtained in \citet{Nogueras-Lara:2020aa}. This is in good agreement with the value $A_{K_s}=1.82\pm0.14$\,mag obtained in Table\,\ref{prop_cen} for the first extinction layer. This lower extinction, which only reaches the first extinction layer as defined in this paper, is not surprising given the limiting magnitude when using aperture photometry from the VVV survey and RC stars to compute the extinction in the NSD \citep[see Fig.\,12 in ][]{Nogueras-Lara:2019aa}.

Dedicated high-angular-resolution observations for particular GC regions make the situation different for small interesting regions like the nuclear star cluster. In this way, using NACO (NAOS-CONICA) data, \citet{Schodel:2010fk} obtained a high resolution extinction map, $\sim1-2''$, for a region with a radius $\sim20''$ centred on Sagittarius\,A*. The mean extinction towards the central parsec around Sgr\,A* was estimated to be $A_{K_s}=2.54\pm0.12$\,mag \citep{Schodel:2010fk}. To compare this value with our results, we considered stars in a radius of $\sim 13''$ and estimated their extinction combining the first and the second extinction layers defined in this paper. We obtained a mean value of $A_{K_s}=2.40\pm0.28$\,mag, in good agreement with the previous value.

\subsection{Comparison with 3D extinction maps}

We aimed to compare our extinction maps with more general 3D extinction maps, in spite of their different resolutions and depths. In this way, this comparison is limited by their typical angular resolution of several arcminutes and by the sensitivities of the surveys used  \citep[see Sect. 1 in][]{Schultheis:2009tg}. This implies that some 3D extinction maps, like the one derived by \citet{Marshall:2006ty}, are limited by relatively low extinctions ($A_v\sim25$\,mag) and by the underlying models used to build them \citep[e.g.][]{Drimmel:2003vv}. Therefore, they cannot be directly compared with our results. 

On the other hand, we used the more recent 3D extinction map obtained by \citet{Chen:2013vm}, whose extinction values for the inner bulge and the central region appear to be in agreement with deeper 2D studies \citep[see Sect. 5.3 in][]{Chen:2013vm}. The extinction map is presented at a resolution of $15'\times15'$ for six different combinations of colours in distance bins of 1\,kpc. We chose the lines of sight towards the Central region of the GNS survey and computed average values of the colour excesses $E(J-K)=3.99\pm0.23$\,mag and $E(J-K)=1.34\pm0.12$\,mag, at the GC distance of $\sim8$\,kpc. The uncertainties correspond to the standard deviation of the values obtained for the different lines of sight. Using the extinction curve obtained by \citet{Nogueras-Lara:2020aa}, we transformed the colour excesses and obtained $A_{K_s\ JK_s} = 1.64\pm0.09$\,mag, and $A_{K_s\ HK_s} = 1.59\pm0.14$\,mag. The subindices indicate the used colour excess. These values are in agreement with $A_{K_s}=1.82\pm0.14$ obtained in Table\,\ref{prop_cen} for the first layer of the Central GNS region, where the uncertainty corresponds to the standard deviation of the extinction values. We adopted the first layer for comparison due to the limitations imposed by the Two Micron All Sky Survey (2MASS) and the Galactic Legacy Infrared Midplane Survey Extraordinaire II (GLIMPSE-II) used to compute the 3D extinction maps.

\subsection{Previous work using the GNS survey}
    
We also compared our results with previous work carried out using the GNS survey, which calculated average extinction values for different regions:

(1) \citet{Nogueras-Lara:2018aa} obtained $A_{K_s}=1.86\pm0.10$\,mag for pointing \#1 of the GNS survey \citep[see Fig. 2 of ][]{Nogueras-Lara:2019aa}. The value was computed using stars detected in $JHK_s$ (analogous to the first extinction layer presented here) and the extinction index $\alpha_{JHK_s} = 2.30\pm0.08$ \citep{Nogueras-Lara:2018aa}. 

(2) \citet{Nogueras-Lara:2018ab} computed $A_{K_s}=1.14\pm0.09$\,mag and $A_{K_s}=1.39\pm0.09$\,mag for pointings \#B1 and \#B2 from the inner bulge regions I B N and I B S, respectively, where the uncertainties correspond to the systematics.

We found that the average values agree well within the uncertainties with the extinction values presented in Table \ref{prop_cen} for the Central region ($\bar{A}_{K_s}=1.82\pm0.12$ mag), the IBN ($\bar{A}_{K_s}=1.18\pm0.09$ mag), and the IBS ($\bar{A}_{K_s}=1.49\pm0.10$ mag). The uncertainties were obtained propagating the statistical and systematic uncertainties of the average extinction values.    
    
 \section{De-reddening}

We de-reddened the GNS photometry using the two-layers approach described previously, and obtained a catalogue of de-reddened sources.

\subsection{Results}

We excluded the foreground population and assigned each star to its corresponding extinction layer following the criterion explained in Sect. \ref{sel_criterion}, as indicated in Fig.\,\ref{fig_star_sel}. For each band and layer, we obtained de-reddened photometry independently applying the obtained extinction maps.

For stars in the first extinction layer, we applied the extinction maps $A_{J,\ JH}$ and $A_{H,\ JH}$, (obtained using $JH$ bands, as the subindices indicate), and $A_{H,\ HK_s}$ and $A_{K_s,\ HK_s}$, (obtained using $HK_s$ bands). The result for each star is an associated de-reddened value for $J$ and $K_s$ photometry, and two de-reddened values for the $H$ band, corresponding to the extinction maps computed using either $JH$ or $HK_s$ photometry.

For stars in the second layer, we applied the second layer extinction maps $A_H$ and $A_{K_s}$, and computed the de-reddened photometry for the $H$ and $K_s$ bands.

\begin{figure*}
   \includegraphics[width=\linewidth]{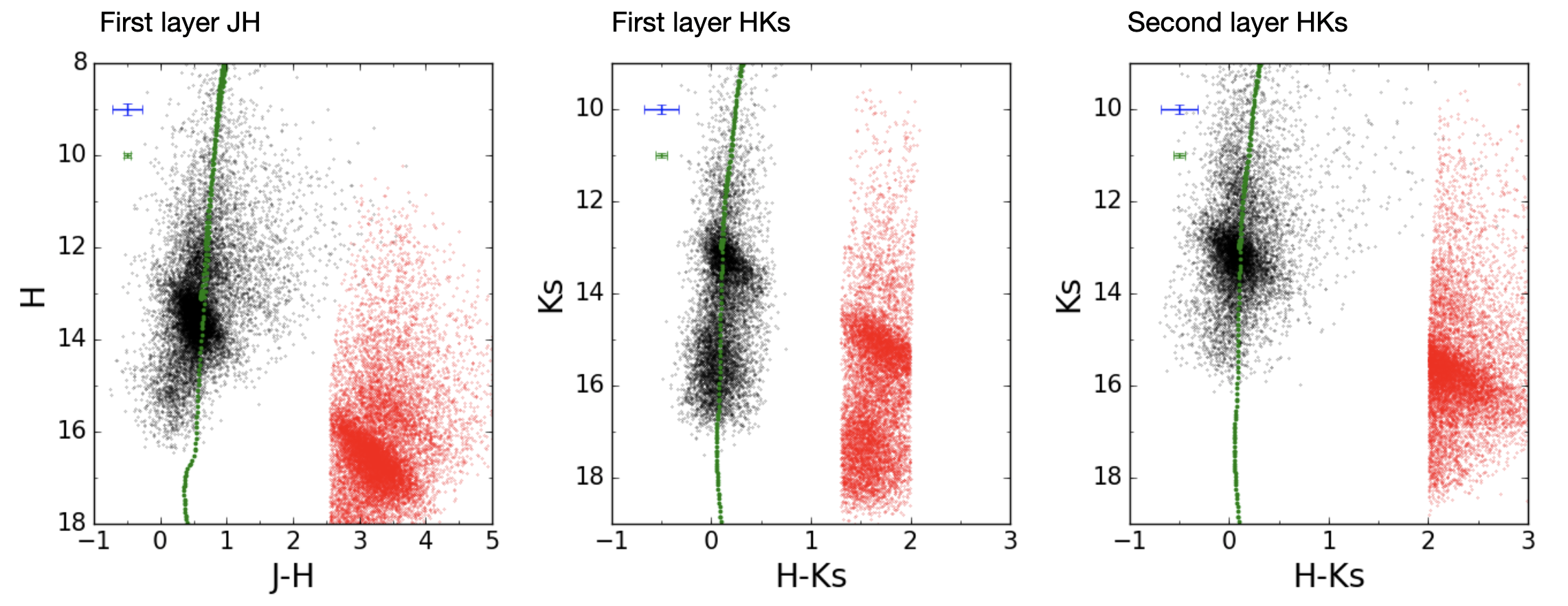}
   \caption{Colour-magnitude diagrams before and after the application of the extinction maps to the Central region of the GNS survey. The red and black dots correspond to the original and the de-reddened CMDs, respectively. Only a fraction of stars are plotted given the high number of sources. The labels above each panel indicate the corresponding extinction layer. The green line depicts a Parsec isochrone of $\sim 10$ Gyr with twice solar metallicity. The blue and green error bars indicate the systematic uncertainties of the de-reddening and the ZP, respectively.}

   \label{Central_dereddened}
    \end{figure*}

The results obtained for the Central region are shown in Fig.\,\ref{Central_dereddened}. The spread of the data points and the width of the standard distribution of the RC sources is  $\sim2$ times lower in the corrected diagrams. Therefore, the differential extinction is significantly corrected in all the cases. 

We over-plotted a Parsec isochrone \citep[release v1.2S + COLIBRI S\_35 + PR16, ][]{Bressan:2012aa,Chen:2014aa,Chen:2015aa,Tang:2014rm,Marigo:2017aa,Pastorelli:2019aa} of an old ($\sim 10$\,Gyr) metal rich (twice solar) stellar population, which is the dominant stellar population in the NSD \citep{Nogueras-Lara:2019ad}, to check that the de-reddening was appropriate. We observed a deviation of the de-reddened data from the model at the faint end ($K_s\sim15-16$\,mag) due to the incompleteness of the data \citep{Nogueras-Lara:2020aa}. The de-reddened CMDs corresponding to the remaining regions are shown in  Appendix\,\ref{dereddenedCMDs} (Figs. \ref{dereddened1} and \ref{dereddened2}).

\subsection{Catalogue}
 
We created a new version of the GNS catalogue by adding the results obtained  for extinction. We present an independent catalogue for each of the regions of the survey in the same way as we did in \citet{Nogueras-Lara:2019aa}. We followed the scheme shown in Table 2 of \citet{Nogueras-Lara:2019aa} and used the new catalogue corrected for duplications (see Sect.\,\ref{dupli}). For simplicity, we only included the absolute coordinates of each source ($RA$, $\Delta RA$, $DEC$, and $\Delta DEC$) and omitted the coordinates corresponding to the detection of each source in any given band (they are available in the original catalogue corrected for duplications). We also included the photometry and the corresponding uncertainties for all three bands. A value of 99 for the photometry indicates non-detection for a given source in a given band.

We added two columns indicating whether a star is detected as foreground population using the methodology explained in Sect. \ref{sel_criterion}. Since the identification was carried out using the CMDs $H$ versus $J-H$ and $K_s$ versus $H-K_s$, we specified the detection as foreground population for a given source in these two diagrams using $F_{JH}$ and $F_{HK_s}$, respectively. A value of 1 indicates that a source is detected as foreground population, whereas a value of -1 indicates that the source corresponds to the target population (GC stars for Central, NSD E, NSD W, T E, and T W, and inner bulge population for I B S and I B N).  Due to the scatter associated to the photometric uncertainties and the degeneracies between different populations because of  differential extinction, it is unavoidable that the results for some sources disagree between $F_{JH}$ and $F_{HK_s}$. This means that some sources might be catalogued as foreground using $F_{JH}$, but are not using $F_{HK_s}$ and vice versa.

We also added six columns corresponding to the extinction values for each band and layer ($A^i_{band\ diagram }$), and six columns with the associated uncertainties ($A^i_{band\ diagram }$).  The super-index $i$ refers to the extinction layer, whereas the subindex $band\ diagram$ gives information about the band of the extinction maps and the photometric bands used to compute the extinction maps from the reference stars. As it happens to the foreground population, it might occur that some stars are detected as belonging to the first layer using a given reference diagram and to the second one using the other. For instance, a star might belong to the first layer using $JH$ as a reference, but be identified as from the second one when using $HK_s$. This is due to the intrinsic scatter of the GC population and cannot be avoided. A value of -1 indicates that a source does not belong to a given extinction layer. A value of 0 indicates that a source is too close to the edge of the corresponding extinction map and thus the extinction value was not computed. A value of 'nan' indicates that there is no associated value for the extinction map given the low number of reference stars.

The two-layers approach was not necessary in the case of the inner bulge regions (I B S and I B N), as  was explained previously (see Sect. \ref{sel_criterion} for further details). Therefore, only one layer is specified (four columns for extinction and four for its associated uncertainties).

Table \ref{final_cat} shows part of the catalogue obtained for the Central region as an example. All the catalogues presented in this paper are made publicly available at the CDS.

\setlength{\tabcolsep}{2.35pt}
\def\arraystretch{1.5}
       \begin{sidewaystable*}
       \footnotesize
       
\begin{center}
\caption{Fields included in each catalogue.}
\label{final_cat}

\begin{tabular}{cccccccccccccccccccccccc}
 &  &  &  &  &  &  &  &  &  &  &  &  &  &  &  &  &  &  &  &  &  &  & \tabularnewline
\hline 
\hline 
RA &  $\Delta$RA & DEC & $\Delta$DEC & $J$ & d$J$ & $H$ & d$H$ & $K_s$ & d$K_s$ & $F_{JH}$ & $F_{HK_s}$ &$ A^1_J$ $_{JH}$ & d$A^1_J$ $_{JH}$ & $ A^1_H$ $_{JH}$  & d$A^1_H$ $_{JH}$ & $A^1_H$ $_{HK_s}$ & d$A^1_H$ $_{HK_s}$ & $A^1_{K_s}$ $_{HK_s}$ & d$A^1_{K_s}$ $_{HK_s}$ & $ A^2_H$ $_{HK_s}$ & d$A^2_H$ $_{HK_s}$ & $A^2_{K_s}$ $_{HK_s}$ & d$A^2_{K_s}$ $_{HK_s}$\tabularnewline
$^\circ$  & $''$ & $^\circ$  & $''$ & (mag) & (mag) &  (mag)& (mag) & (mag) & (mag) &  &  & (mag) & (mag) & (mag) & (mag) & (mag)& (mag) & (mag) & (mag) & (mag) & (mag) & (mag) & (mag) \tabularnewline
\hline 
266.48 & 0.008 & -28.97 & 0.006 & 19.110 & 0.011 & 15.419 & 0.007 & 13.568 & 0.006 & -1 & -1 & 5.886 & 0.312 & 3.147 & 0.167 & 3.555 & 0.119 & 1.932 & 0.065 & -1.000 & -1.000 & -1.000 & -1.000\tabularnewline
266.47 & 0.007 & -28.97 & 0.008 & 19.093 & 0.011 & 18.090 & 0.026 & 16.732 & 0.013 & 1 & -1 & -1.000 & -1.000 & -1.000 & -1.000 & 3.517 & 0.117 & 1.912 & 0.063 & -1.000 & -1.000 & -1.000 & -1.000\tabularnewline
266.48 & 0.001 & -28.97 & 0.001 & 19.139 & 0.019 & 14.994 & 0.010 & 12.848 & 0.006 & -1 & -1 & nan & nan & nan & nan & -1.000 & -1.000 & -1.000 & -1.000 & 5.875 & 0.458 & 3.193 & 0.249\tabularnewline
266.47 & 0.002 & -28.96 & 0.002 & 19.161 & 0.014 & 17.189 & 0.016 & 16.230 & 0.006 & 1 & 1 & -1.000 & -1.000 & -1.000 & -1.000 & -1.000 & -1.000 & -1.000 & -1.000 & -1.000 & -1.000 & -1.000 & -1.000\tabularnewline
266.47 & 0.008 & -28.97 & 0.008 & 19.171 & 0.035 & 15.830 & 0.009 & 14.112 & 0.006 & -1 & -1 & 6.177 & 0.084 & 3.303 & 0.045 & 3.744 & 0.101 & 2.035 & 0.055 & -1.000 & -1.000 & -1.000 & -1.000\tabularnewline
266.48 & 0.007 & -28.97 & 0.006 & 19.151 & 0.017 & 17.126 & 0.009 & 16.119 & 0.010 & 1 & 1 & -1.000 & -1.000 & -1.000 & -1.000 & -1.000 & -1.000 & -1.000 & -1.000 & -1.000 & -1.000 & -1.000 & -1.000\tabularnewline
266.47 & 0.002 & -28.97 & 0.002 & 19.183 & 0.011 & 15.912 & 0.007 & 14.267 & 0.004 & -1 & -1 & 5.815 & 0.229 & 3.110 & 0.123 & 3.539 & 0.092 & 1.923 & 0.050 & -1.000 & -1.000 & -1.000 & -1.000\tabularnewline
266.48 & 0.007 & -28.97 & 0.008 & 19.266 & 0.019 & 18.190 & 0.018 & 17.419 & 0.016 & 1 & 1 & -1.000 & -1.000 & -1.000 & -1.000 & -1.000 & -1.000 & -1.000 & -1.000 & -1.000 & -1.000 & -1.000 & -1.000\tabularnewline
266.48 & 0.008 & -28.97 & 0.006 & 19.156 & 0.014 & 15.524 & 0.008 & 13.686 & 0.006 & -1 & -1 & 6.057 & 0.311 & 3.239 & 0.166 & 3.644 & 0.065 & 1.980 & 0.035 & -1.000 & -1.000 & -1.000 & -1.000\tabularnewline
266.48 & 0.003 & -28.98 & 0.003 & 19.276 & 0.018 & 17.000 & 0.010 & 15.752 & 0.009 & 1 & 1 & -1.000 & -1.000 & -1.000 & -1.000 & -1.000 & -1.000 & -1.000 & -1.000 & -1.000 & -1.000 & -1.000 & -1.000\tabularnewline
266.47 & 0.002 & -28.97 & 0.002 & 19.232 & 0.009 & 14.402 & 0.008 & 11.812 & 0.005 & -1 & -1 & 6.038 & 0.168 & 3.229 & 0.090 & -1.000 & -1.000 & -1.000 & -1.000 & 5.455 & 0.364 & 2.965 & 0.198\tabularnewline
266.48 & 0.004 & -28.97 & 0.005 & 19.261 & 0.011 & 17.338 & 0.011 & 16.372 & 0.009 & 1 & 1 & -1.000 & -1.000 & -1.000 & -1.000 & -1.000 & -1.000 & -1.000 & -1.000 & -1.000 & -1.000 & -1.000 & -1.000\tabularnewline
266.47 & 0.008 & -28.96 & 0.007 & 19.338 & 0.011 & 15.593 & 0.010 & 13.621 & 0.006 & -1 & -1 & 6.072 & 0.287 & 3.247 & 0.154 & 3.688 & 0.145 & 2.004 & 0.079 & -1.000 & -1.000 & -1.000 & -1.000\tabularnewline
266.47 & 0.007 & -28.97 & 0.006 & 19.253 & 0.011 & 15.648 & 0.008 & 13.732 & 0.007 & -1 & -1 & 6.368 & 0.112 & 3.406 & 0.060 & 3.771 & 0.079 & 2.049 & 0.043 & -1.000 & -1.000 & -1.000 & -1.000\tabularnewline
266.47 & 0.002 & -28.96 & 0.002 & 19.235 & 0.016 & 15.610 & 0.008 & 13.784 & 0.005 & -1 & -1 & 6.550 & 0.081 & 3.503 & 0.043 & 3.848 & 0.056 & 2.091 & 0.031 & -1.000 & -1.000 & -1.000 & -1.000\tabularnewline
266.48 & 0.003 & -28.97 & 0.003 & 19.221 & 0.017 & 15.467 & 0.008 & 13.575 & 0.007 & -1 & -1 & 6.645 & 0.154 & 3.554 & 0.083 & 3.850 & 0.087 & 2.092 & 0.047 & -1.000 & -1.000 & -1.000 & -1.000\tabularnewline
266.47 & 0.008 & -28.97 & 0.007 & 19.259 & 0.033 & 15.681 & 0.007 & 13.736 & 0.005 & -1 & -1 & 5.937 & 0.260 & 3.175 & 0.139 & 3.647 & 0.128 & 1.982 & 0.069 & -1.000 & -1.000 & -1.000 & -1.000\tabularnewline
266.47 & 0.014 & -28.97 & 0.012 & 19.318 & 0.014 & 17.427 & 0.017 & 16.530 & 0.010 & 1 & 1 & -1.000 & -1.000 & -1.000 & -1.000 & -1.000 & -1.000 & -1.000 & -1.000 & -1.000 & -1.000 & -1.000 & -1.000\tabularnewline
266.48 & 0.001 & -28.97 & 0.002 & 19.369 & 0.012 & 14.350 & 0.006 & 11.695 & 0.006 & -1 & -1 & 6.642 & 0.177 & 3.552 & 0.094 & -1.000 & -1.000 & -1.000 & -1.000 & 5.172 & 0.118 & 2.811 & 0.064\tabularnewline
266.47 & 0.003 & -28.96 & 0.004 & 19.350 & 0.018 & 17.516 & 0.015 & 16.721 & 0.015 & 1 & 1 & -1.000 & -1.000 & -1.000 & -1.000 & -1.000 & -1.000 & -1.000 & -1.000 & -1.000 & -1.000 & -1.000 & -1.000\tabularnewline
266.47 & 0.010 & -28.97 & 0.006 & 19.403 & 0.012 & 16.013 & 0.006 & 14.270 & 0.006 & -1 & -1 & 6.320 & 0.089 & 3.380 & 0.048 & 3.728 & 0.098 & 2.026 & 0.053 & -1.000 & -1.000 & -1.000 & -1.000\tabularnewline
266.48 & 0.003 & -28.97 & 0.003 & 19.322 & 0.021 & 15.029 & 0.006 & 12.793 & 0.006 & -1 & -1 & 6.309 & 0.159 & 3.374 & 0.085 & -1.000 & -1.000 & -1.000 & -1.000 & 4.877 & 0.090 & 2.650 & 0.049\tabularnewline
... & ... & ... & ... & ... & ... & ... & ... & ... & ... & ... & ... & ... & ... & ... & ... & ... & ... & ... & ... & ... & ... & ... & ...\tabularnewline
\hline 
\end{tabular}

\end{center}

\footnotesize
\textbf{Notes.} Example of the final catalogue. Rows 120-142 correspond to the Central region of the GNS survey. The position and photometry of the sources were taken from the original catalogue \citep[see Table 2 of][]{Nogueras-Lara:2019aa}. The right ascension and the declination
of the sources are RA, $\Delta$RA,   DEC, and $\Delta$DEC. The coordinates were rounded to two decimals in this table. The photometry and the associated uncertainties are indicated by $J$, $dJ$, $H$, $dH$, $K_s$ , and $dK_s$. A 99 indicates that there is no detection for a given source. $F_{i}$ indicates whether a source is detected as foreground population (=1) using the bands specified as a reference ($i$).  A value of -1 refers to a non-foreground population. $A^i_{j}$ and $dA^i_{j}$ correspond to the value of the extinction and its associated uncertainty. The subindices $i$ and $j$ are the bands used to compute the extinction maps and the extinction layer, respectively. A value of -1 indicates that a star does not correspond to a given layer. A value of 0 indicates a star too close to the borders of the corresponding extinction map. A 'nan' value indicates that there is no  associated value.

\end{sidewaystable*}

\section{Limitations of the de-reddening}

We studied the limitations of the de-reddening process, giving the adopted strategy, and its application to different stellar types analysing possible non-linear effects.

\subsection{Extinction layers}

Extinction is a 3D phenomenon that needs to be simplified to be taken into account in an affordable way. The reference stars used to create the extinction maps are distributed along the line of sight. Although the distance does not affect the measured stellar colours directly, the amount of reddening material towards the stars is a function of their distance. We used the two-layers approach, dividing the stars by means of a colour cut in the CMDs. In this way, we avoided mixing reference stars whose absolute extinctions are too different. This approach can be generalised to be applied to the whole survey, as explained in previous sections. It gives reasonably good average values for the whole stellar population (see Table\,\ref{prop_cen}). Nevertheless, it also produces over- or under- corrections for some of the stars, as shown by the scatter of the de-reddened CMDs. To correct specific regions where the differential extinction is too large, therefore, a more convenient approach would be to impose constraints on the colour range to accept reference stars for a given extinction-map pixel in order to avoid mixing stars with too different extinctions (see Methods section in \citealt{Nogueras-Lara:2019ad}).

The extinction layers that we present here do not necessarily correspond to independent physical layers. They are the result of the incompleteness of the $J$ band data and serve as a boundary to avoid computing the extinction maps using stars that are  close in projection but suffer from very different extinctions. We recommend therefore that de-reddened stars from independent layers should not be mixed in order to avoid possible systematics related to the cuts introduced when defining the layers. In those cases, the best approach is to create  dedicated extinction maps following the approach described in \citet{Nogueras-Lara:2019ad}, imposing constraints on the colour differences for the neighbouring stars to avoid too different extinctions.

The GC is a complex and clumpy region characterised by the presence of dark clouds that produce strong extinction variations on arc-second scales \citep[e.g.][]{Launhardt:2002nx,Nogueras-Lara:2018aa,Battersby:2020vt}. To account for these variations, we combined a number of reference stars per pixel, restricting the distance from a given pixel to the reference stars and introducing the IDW method to weight the stars depending on their distance to a given pixel. Moreover, the two-layers approach helps us to avoid the combination of stars with very different extinctions but close in projection.

\subsection{Non-linear effects}

There are some non-linear effects, also known as bandwidth effects, which might affect the extinction correction that we present here \citep[][]{1980MNRAS.192..359J,2008BaltA..17..277S,Maiz-Apellaniz:2020aa}.

One of the most relevant effects is that the extinction is not linear with the stellar type. We built extinction maps using RC stars and applied them to  the target stars to correct their photometry. This means that other types of stars, different from the red giants used as a reference, might be affected by the extinction in a slightly different way that is not considered in our correction. To analyse this effect, we computed the extinction corresponding to the $JHK_s$ bands for stars with different temperatures, using the intrinsic and reddened $H-K_s$ colours. We used colours instead of magnitudes because they do not depend on distance and stellar radius. To compute them, we assumed the corresponding Kurucz models \citep{Kurucz:1993fk}, the HAWK-I filter curves \citep{Kissler-Patig:2008uq}, the extinction curve derived in \citet{Nogueras-Lara:2020aa}, and a monochromatic amount of extinction of $A_{1.61} = 3.40$\,mag for all of them \citep{Nogueras-Lara:2020aa}. In the latter, the subindex $1.61$ indicates the monochromatic wavelength 1.61\,$\mu$m and the selected value corresponds to the average extinction to the central region of the GNS survey following Table\,2 in \citet{Nogueras-Lara:2020aa}.

We used Eq. \ref{eq_ratio} to calculate the $A_{K_s}$ values, substituting the intrinsic and the de-reddened colours previously computed. Table \ref{stellar_types} shows the results. We estimated whether the differences between stellar types are significant considering an uncertainty of $\sim10\,\%$ for the intrinsic colours (as for RC stars, see Sect.\,\ref{intrinsic}). We obtained that there is no significant difference within the uncertainties for any of the stellar types considered here in comparison with RC stars. Moreover, we translated the $A_{K_s}$ values into $A_H$ and $A_J$ extinctions using the ratios $A_{H}/A_{K_s} = 1.84\pm0.03$ and $A_{J}/A_{K_s} = 3.44\pm0.08$ (Table\,\ref{stellar_types}). The difference between values is larger for shorter wavelengths but the uncertainties also grow accordingly ($\Delta A_H\sim0.14$\,mag and $\Delta A_J\sim0.28$\,mag) when considering the corresponding extinction ratios. Thus, the effect is not significant within the uncertainties. \\

\begin{table}
\caption{Extinction for different stellar types considering $A_{1.61}=3.40$\,mag.}
\label{stellar_types} 
\begin{center}
\def\arraystretch{1.4}
\setlength{\tabcolsep}{3.8pt}

\begin{tabular}{ccccccc}
 &  &  & & & &\tabularnewline
\hline 
\hline 
Stellar  & Temp. & $(H-K_s)_0$ & $H-K_s$ & $A_{K_s}$& $A_{H}$ & $A_{J}$\tabularnewline
type & K & mag & mag & mag & mag & mag\tabularnewline
\hline 
M2I & 3500 & 0.29 & 1.83 & 1.83& 3.37&6.30\tabularnewline
RC & 4750 & 0.09 & 1.65 & 1.86& 3.42&6.40\tabularnewline
A0V & 9500 & -0.01 & 1.57 & 1.88& 3.46&6.47\tabularnewline
BOV & 30000 & -0.09 & 1.47 & 1.86& 3.40&6.36\tabularnewline
O5V & 45000 & -0.13 & 1.45 & 1.88& 3.46&6.47\tabularnewline
\hline 
\end{tabular}

\end{center}
\footnotesize
\textbf{Notes.} The associated uncertainties are $\Delta A_{K_s}\sim$0.07, $\Delta A_{H}\sim$0.14, and $\Delta A_{J}\sim$0.28\,mag. 

 \end{table}

A further small side effect is the influence of the effective wavelength when using wide-band filters to produce the extinction maps. The ratios between $A_H/A_{K_s}$ and $A_J/A_{K_s}$ that we used in this paper were computed using RC stars as a reference \citep[see][]{Nogueras-Lara:2019ac,Nogueras-Lara:2020aa}. In reality, these quantities would vary for each star depending on its stellar type. To estimate the variation, we computed the effective wavelength for each of the stellar types in Table\,\ref{stellar_types} using the method describe in \citet{Nogueras-Lara:2018aa}. We then used the equation $A_{\lambda_1}/A_{\lambda_2} = (\lambda_1/\lambda_2)^{-\alpha_{\lambda_1\lambda_2}}$ \citep{Nogueras-Lara:2020aa} to estimate the influence of the effective wavelength variation on the extinction ratios. Averaging over the values obtained for the used stellar types, we ended up with $A_J/A_H=1.88\pm0.01$ and $A_H/A_{K_s}=1.84\pm0.01$, where the uncertainties refer to the standard deviation of the distributions. The results fully agree with the extinction ratios derived in \citet{Nogueras-Lara:2020aa}. Therefore, we concluded that the effect of the effective wavelength on the de-reddening is not significant within the uncertainties.

\section{De-reddened $K_s$ luminosity functions}

\begin{figure}
   \includegraphics[width=\linewidth]{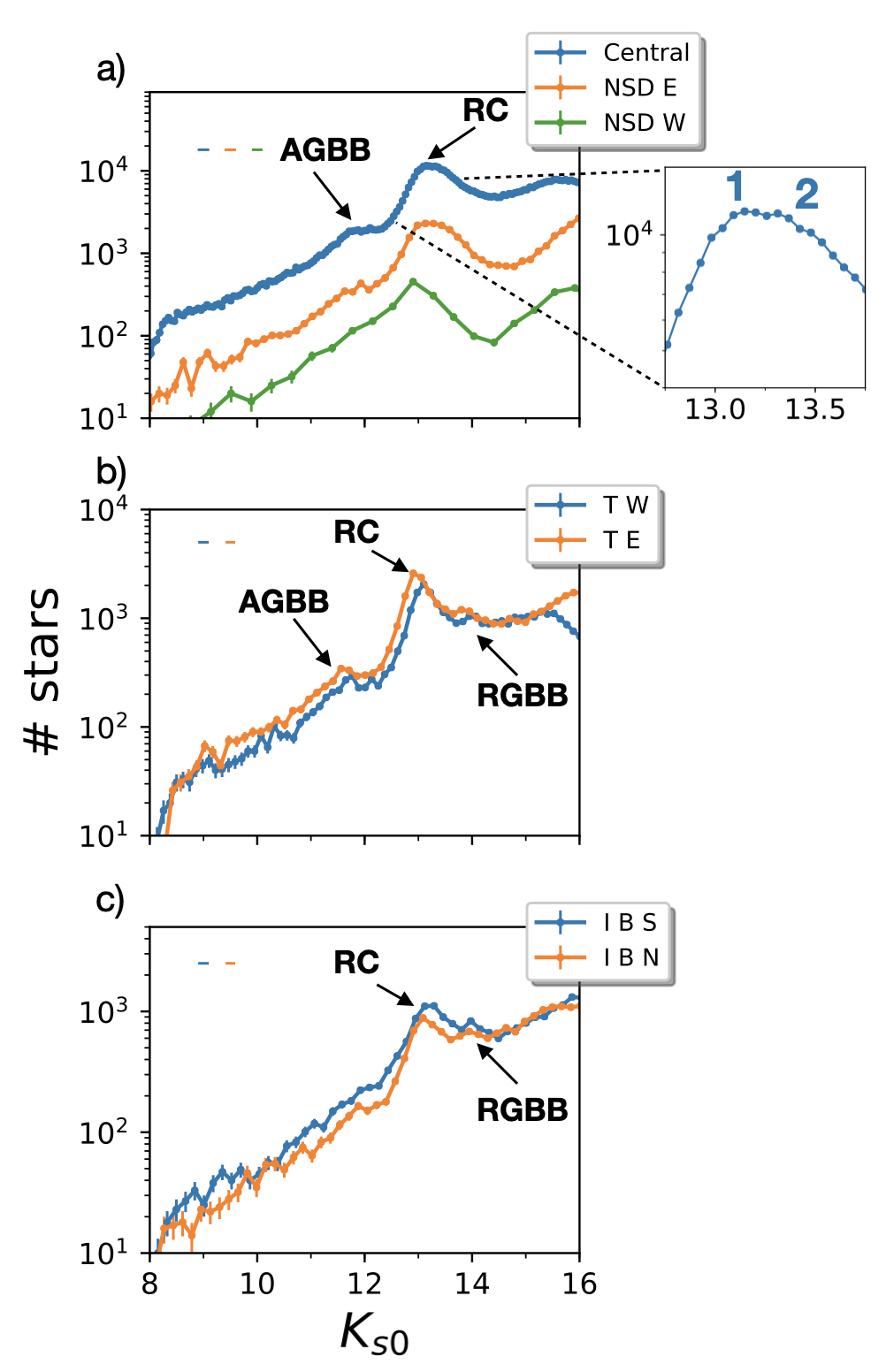}
   \caption{De-reddened KLFs for each of the GNS regions, as indicated in the legend. a) NSD regions. b) Transition regions. c) Inner bulge regions. The uncertainties are estimated as the square root of the number of stars per bin. The bins are different for each case and were computed using the python function numpy.histogram\_bin\_edges \citep{Harris:2020aa}. The RC, AGBB, and the RGBB  are indicated in the panels, where the features are detected. The coloured error bars on the top of the panels indicate the systematic uncertainty due to the de-reddening process and the ZP systematics. The green KLF corresponding to the NSD W suffers from low completeness due to lower than average data quality. The zoom-in panel indicates the double RC feature associated with the Central region.}

   \label{KLFs}
    \end{figure}

To check the usefulness of the extinction maps and the de-reddened  catalogues, we performed a preliminary analysis by building KLFs for each of the regions in the GNS catalogue. We restricted the analysis to stars from the first extinction layer ($A^1_{K_s,\ HK_s}>0$, as indicated by the super-index), where the completeness is higher than in the secondary layer, and the foreground population is excluded (see Fig.\,\ref{fig_star_sel}). We obtained de-reddened KLFs with a completeness larger than 80\,\% at $K_{s0}\sim14.5$\,mag \citep[according to the completeness $\sim80$\,\% at $K_s\sim16.3$\,mag, computed for the Central region in][]{Nogueras-Lara:2020aa}. Only the KLF from the NSD\,W shows lower completeness and a significantly lower number of detected stars due to lower than average data quality for this region \citep[see Table A.3. in][]{Nogueras-Lara:2019aa}. Figure\,\ref{KLFs} shows the corresponding KLFs. The regions used to produce the KLFs are larger than those used in previous studies by a factor of $\sim3$ for the NSD regions \citep{Nogueras-Lara:2019ad} and $\sim2$ for the innermost bulge \citep{Nogueras-Lara:2018ab}. The transition regions had not been previously studied using the KLFs. 

The main features appearing in the KLFs correspond to the asymptotic giant branch bump (AGBB), the RC, and the red giant branch bump (RGBB) \citep[see Fig.\,2 in supplementary material of][]{Nogueras-Lara:2019ad}. Analysing the relative fraction of stars between them, it is possible to infer the star formation history (SFH) of the regions \citep[e.g.][]{Nogueras-Lara:2019ad}. 

A preliminary analysis indicates significant differences between the KLFs corresponding to the different regions studied. We observed that the shape of the KLFs varies with latitude, being similar for regions within the same structure. Namely, the KLFs corresponding to regions in the NSD, the inner bulge, and the transition region between the NSD and the bulge present similar features. This points towards physically different components with different SFHs, in agreement with previous work \citep[e.g.][]{Launhardt:2002nx,Nishiyama:2013uq,Nogueras-Lara:2018ab,gallego-cano2019,Nogueras-Lara:2019ad}. In particular, the RC feature appears to be thicker for the NSD regions (excluding the NSD W, whose lower completeness avoids a good coverage for faint $K_s$ magnitudes), which is probably the result of the presence of a secondary RC feature due to a star formation event $\sim1$\,Gyr ago that was responsible for the formation of $\sim5$\,\% of the stellar mass \citep{Nogueras-Lara:2019ad}.%This double RC feature appears to be  more pronounced in the case of the NSD E, so maybe the presence of RC stars with ages $\sim1$\,Gyr is more significant in this region. 

The double RC feature does not appear in the case of the inner bulge regions, where the RGBB feature is more prominent (and the AGBB less important), indicating a different SFH dominated by mainly old stars according to previous work on the inner bulge \citep{Nogueras-Lara:2018ab}. Moreover, the AGBB is more prominent in the NSD regions, indicating a larger contribution of young stars (several hundred million years) to the KLF. On the other hand, the KLFs corresponding to the transition regions share characteristics from both, the NSD and the inner bulge fields, though are more influenced by the latter. Thus, this region appears as a border between these two different components of the innermost region of the Galaxy.

\section{Conclusions}
    
In this paper we present high-angular-resolution extinction maps ($\sim3''$) for the whole GNS catalogue and de-redden the GNS photometry by discriminating the foreground population. We make the maps and the de-reddened catalogues publicly available to the community. 

We identified duplicated sources in the original GNS catalogue and corrected them to produce a clean second data release. We created CMDs with the new data and identified the foreground stellar population via colour cuts in the CMDs $H$ versus $J-H$ and $K_s$ versus $H-K_s$. Given that the $J$ band is more affected by extinction than the $H$ and $K_s$ bands, we defined a two-layer approach to build extinction maps to de-redden the data in a consistent way. We used RC and red giant stars of similar brightness as a reference for the extinction maps and computed mean intrinsic colours $(J-H)_0=0.52\pm0.04$ and $(H-K_s)_0=0.10\pm0.01$, averaging over previous values and computing new ones using a Parsec synthetic stellar population. We obtained mean values for the extinction ($A_J$, $A_H$, and $A_{K_s}$) for all the regions covered by the GNS, de-reddening the stars belonging to each of the two extinction layers considered (Table\,\ref{prop_cen}). 

To check the quality of the extinction maps, we compared our results with previous values for the analysed regions and found a good agreement between them. Moreover, we also assessed the differential extinction correction and computed de-reddened CMDs. We found that the scatter of the RC features is  significantly reduced (a factor $\sim 2$) after applying the extinction maps. Finally, we estimated the statistical and systematic uncertainties for the extinction maps and ended up with values $\lesssim 5$\,\% of the mean extinction in both cases. 

The presented extinction maps supersede previous ones for the innermost regions of the Galaxy because they used the GNS survey, which constitutes the highest angular resolution photometric survey in the NIR for the innermost regions of the Milky Way \citep{Nogueras-Lara:2019aa}. They also comprise the whole GNS survey covering a region ($\sim 6000$\,pc$^2$), which is approximately four times larger than previous extinction maps using the GNS survey.  

Given that we computed two $A_H$ extinction maps for the first extinction layer (using $JH$ and $HK_s$ photometry), we assessed the  extinction curve used \citep{Nogueras-Lara:2019ac, Nogueras-Lara:2020aa} and compared the mean values obtained for the extinction $A_H$ for each of the extinction maps. They agree within the uncertainties, being consistent with the extinction curve given by the ratios $A_J/A_H = 1.87\pm0.03$ and $A_H/A_{K_s}= 1.84\pm0.03$.

We discussed the limitations of the de-reddening process using the computed extinction maps and, paying special attention to the complexity of translating a 3D process into two 2D extinction layers, and we analysed the possible over- or under- corrections for individual stars. We also discussed non-linear effects affecting the extinction maps, such as the de-reddening of different stellar types and the influence of the effective wavelength when using broad-band filters. We find that these effects are negligible given the uncertainties of the de-reddening process.

To check the effectiveness of the extinction correction, we compared the extinction-map approach with de-reddening on a star-by-star basis (Appendix\,\ref{app}). We did this by simulating a stellar population according to the one expected for the NSD \citep{Nogueras-Lara:2019ad}, also assuming  the presence of variable stars. We concluded that the extinction-map approach is more appropriate to de-redden the stars. In particular, the performance of the extinction maps is significantly better when considering the influence of variable stars. Moreover, the extinction maps using RC stars  suffer less from systematics effects associated to the a-priori unknown intrinsic colours of each of the de-reddened stars. On the other hand, the extinction maps also allow us to improve the correction for 3D effects because they used more than a single star close in projection.

Finally, we used the de-reddened stars from the first extinction layer to build KLFs and compared the stellar populations and SFHs from the different regions covered by the GNS survey. We concluded that the regions corresponding to the NSD present a significantly different KLF in comparison with the inner bulge regions, pointing towards different stellar populations and SFHs, in agreement with previous work \citep{Launhardt:2002nx,Nishiyama:2013uq,Nogueras-Lara:2018ab,gallego-cano2019,Nogueras-Lara:2019ad}.

  \begin{acknowledgements}
      This work is based on observations made with ESO
      Telescopes at the La Silla Paranal Observatory under program
      ID 195.B-0283. We thank the staff of
      ESO for their great efforts and helpfulness. F.N.-L. acknowledges the sponsorship provided by the Federal Ministry for Education and Research of Germany through the Alexander von Humboldt Foundation. F.N.-L. and N.N. gratefully acknowledge support by the Deutsche Forschungsgemeinschaft (DFG, German Research Foundation) – Project-ID 138713538 – SFB 881 ("The Milky Way System", subproject B8). R.S. acknowledges financial support from the State
Agency for Research of the Spanish MCIU through the "Center of Excellence Severo
Ochoa" award for the Instituto de Astrof\'isica de Andaluc\'ia (SEV-2017-0709). R.S.  acknowledges financial support from national project
PGC2018-095049-B-C21 (MCIU/AEI/FEDER, UE).
\end{acknowledgements}

\bibliographystyle{aa}
\bibliography{../../BibGC}
%\bibliography{/Users/rainer/Documents/BibDesk/BibGC}

\appendix

\section{De-reddening methods}
\label{app}

Extinction can be determined using a star-by-star approach \citep[e.g.][]{Fritz:2020aa} or building extinction maps averaging over the values of several stars close in projection \citep[e.g.][]{Schodel:2010fk,Nogueras-Lara:2018aa,Nogueras-Lara:2019ad,Fritz:2020aa}. To analyse both methods and to obtain the best performance for the GC, we created a synthetic stellar population of known extinction to apply both techniques and compare the results using $H$ and $K_s$ photometry.

\subsection{Synthetic model}
\label{synt}

We created a synthetic stellar population using Parsec models \citep[release v1.2S + COLIBRI S\_35 + PR16, ][]{Bressan:2012aa,Chen:2014aa,Chen:2015aa,Tang:2014rm,Marigo:2017aa,Pastorelli:2019aa} and assuming the star formation history derived in \citet{Nogueras-Lara:2019ad} for the NSD. Namely, we assumed single age stellar populations of 12, 1.5, 0.25, 0.1, and 0.05 Gyr, accounting respectively for $91\,\%, 5\,\%, 2\,\%, 1\,\%$, and $1\,\%$ of the total mass and twice solar metallicity \citep[e.g.][]{Feldmeier-Krause:2017kq,Schultheis:2019aa,Nogueras-Lara:2019ad}. 

We accounted for variability considering two simple toy models corresponding to two different scenarios. The first one assumes that $\sim15\,\%$ of the stars are variable following the results of \citet{2017MNRAS.470.3427D} for a $2.3'\times2.3'$ region centred on Sagittarius A*. The second one is a more extreme case, $\sim50\,\%$ of variable stars, based on the results obtained by \citet{Gautam:2019aa} for the innermost region of the NSC ($\sim$0.4\,pc). We assumed a mean amplitude of $H=0.25$\,mag for the variable stars \citep[see Fig. 9 in][]{2017MNRAS.470.3427D}. To account for the variability in $K_s$, we transformed the mean amplitude using Eq.\,B5 in \citet{Feast:2008aa}. This equation was derived for RR Lyraes but given that we considered that the observations for different bands were not taken simultaneously, we did not assume any time correlation between the amplitudes in both bands for each varying star. Thus, it is a way of considering that a variable star in $H$ band is also variable in $K_s$, with an amplitude varying within the limits of the transformation used. In this way, we used two independent Gaussian distributions with standard deviations equal to the mean amplitude for each band to simulate the variability. The described approach is independent of the type of variable stars and does not consider any dependence on the variability period but on the amplitude.

To simulate the extinction and a realistic distribution of the stars following the crowding conditions, we used the extinction map derived in \citet{Nogueras-Lara:2019ad} and the GNS catalogue. We selected a region of $1.3'\times1.3'$ and placed the synthetic stars following the positions of real stars from the survey and the extinctions corresponding to the map. Given that too high extinction values are mainly associated with the faint end of the RC features used to create the original extinction map, and thus might be associated with data incompleteness, we removed them to keep a quasi-Gaussian distribution of the extinctions. The uncertainties for the $H$ and $K_s$ photometry were simulated considering a Gaussian distribution with a standard deviation of 0.05\,mag in agreement with the GNS uncertainties \citep{Nogueras-Lara:2019aa}.

To simulate the distribution of the stars in the NSD, we assumed a Gaussian distribution centred on 8\,kpc \citep[e.g.][]{Gravity-Collaboration:2018aa,Do:2019aa}, with a standard distribution of 150\,pc, corresponding to the effective radius of the NSD \citep[e.g.][]{Nishiyama:2013uq,gallego-cano2019}.

\begin{figure}
   \includegraphics[width=\linewidth]{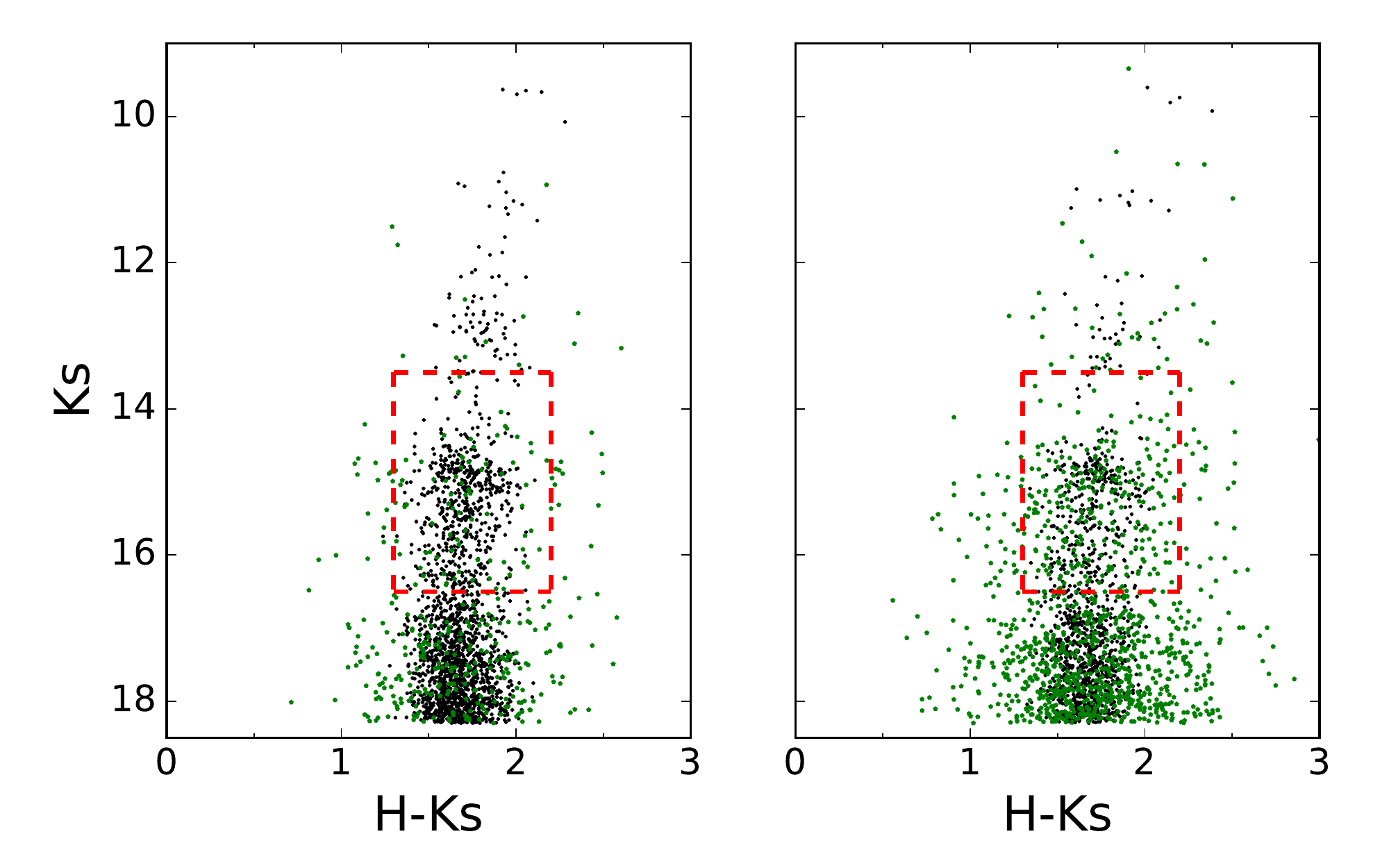}
   \caption{Simulated CMDs $K_s$ vs. $H-K_s$. Left panel: 15\,\% of variable stars. Right panel: 50\,\% of variable stars. The green dots correspond to variable stars. The red dashed lines indicate the selected stars to create the extinction maps.}

   \label{CMD_syn}
    \end{figure}

We kept all the stars with $K_s<18.3$\,mag, corresponding to the 50\,\% completeness limit of the GNS survey \citep{Nogueras-Lara:2020aa}. Figure\,\ref{CMD_syn} shows the obtained CMDs ($K_s$ vs. $H-K_s$) for the considered scenarios of variability.

\subsection{Comparison between methods}
\label{ext_map_method}

We used two methods to de-redden the synthetic photometry:

\begin{itemize}

\item {\bf Star by star approach}. We used the NICE method \citep[e.g.][]{Lada:1994uq,Lombardi:2001fk} to de-redden each star, following the equation

\begin{equation}
\label{eq_ratio}
 A_{i} = \frac{j-i-(j-{i})_0}{(A_j/A_{i})-1}\hspace{0.5cm} ,
\end{equation}

\noindent where $i$ and $j$ refer to two NIR photometric bands, $A_{i}$ is the extinction for the $i$ band, and the subindex 0 indicates intrinsic colour. We used $j = H$ and $i = K_s$ bands and the ratio $A_H/A_{K_s}=1.84$ \citep{Nogueras-Lara:2020aa}.  The intrinsic colour $(H-K_s)_0$ is very similar for different stellar types and, in particular, for the red giant stars \citep[see Fig. 29 in ][]{Nogueras-Lara:2018aa}. Moreover, the majority of stars that can be observed in the GC are red giants, given the SFH of the NSD and the extreme source crowding that impedes the observation of main sequence stars \citep[e.g.][]{Nogueras-Lara:2019ad, Nogueras-Lara:2021uq}. Therefore, we assumed a constant intrinsic colour for all the stars $(H-K_s)_0=0.1$\,mag \citep[see Fig. 29 in ][]{Nogueras-Lara:2018aa}. A more refined assignation of intrinsic colour for early type star would require spectroscopic observations that are normally restricted to small key regions in the GC given the extreme number of sources.\\

\item {\bf Extinction maps}. We applied the approach described in \citet{Nogueras-Lara:2018aa, Nogueras-Lara:2018ab, Nogueras-Lara:2019ad}.  We used RC stars as a reference to compute the extinction, given that they are well distributed across the studied fields. To increase the number of reference stars, we also added red giant stars, whose intrinsic colours are similar to RC stars \citep[see Figs. 33 and 34 of][]{Nogueras-Lara:2018aa}. The extinction for each reference star was computed using Eq.\,\ref{eq_ratio}. We determined the extinction at each point by computing an inverse distance weight (IDW) of the five closest stars within a $\sim7.5''$ radius on a grid with $\sim3''$ spacing, following the equation

\begin{equation}
v = \frac{\sum^n_{i=1}\frac{1}{d^p_i}v_i}{\sum^n_{i=1}\frac{1}{d^p_i}} \hspace{0.5cm},
\end{equation}
\vspace{0.2cm}

\noindent where $v$ is the extinction for a given pixel, $d$ is the distance between the centre of the pixel and each reference star, $v_i$ refer to the extinction computed for each reference star following Eq. \ref{eq_ratio}, and $p$ is the weight factor. We adopted $p=0.25,$ which was found to be the best choice in previous work after exploring the space of $p$ values \citep{Nogueras-Lara:2018aa, Nogueras-Lara:2018ab, Nogueras-Lara:2019ad}. 

We did not assign any extinction value ("nan") if less than five reference stars are within the limiting radius.

\end{itemize}

\begin{figure}
   \includegraphics[width=\linewidth]{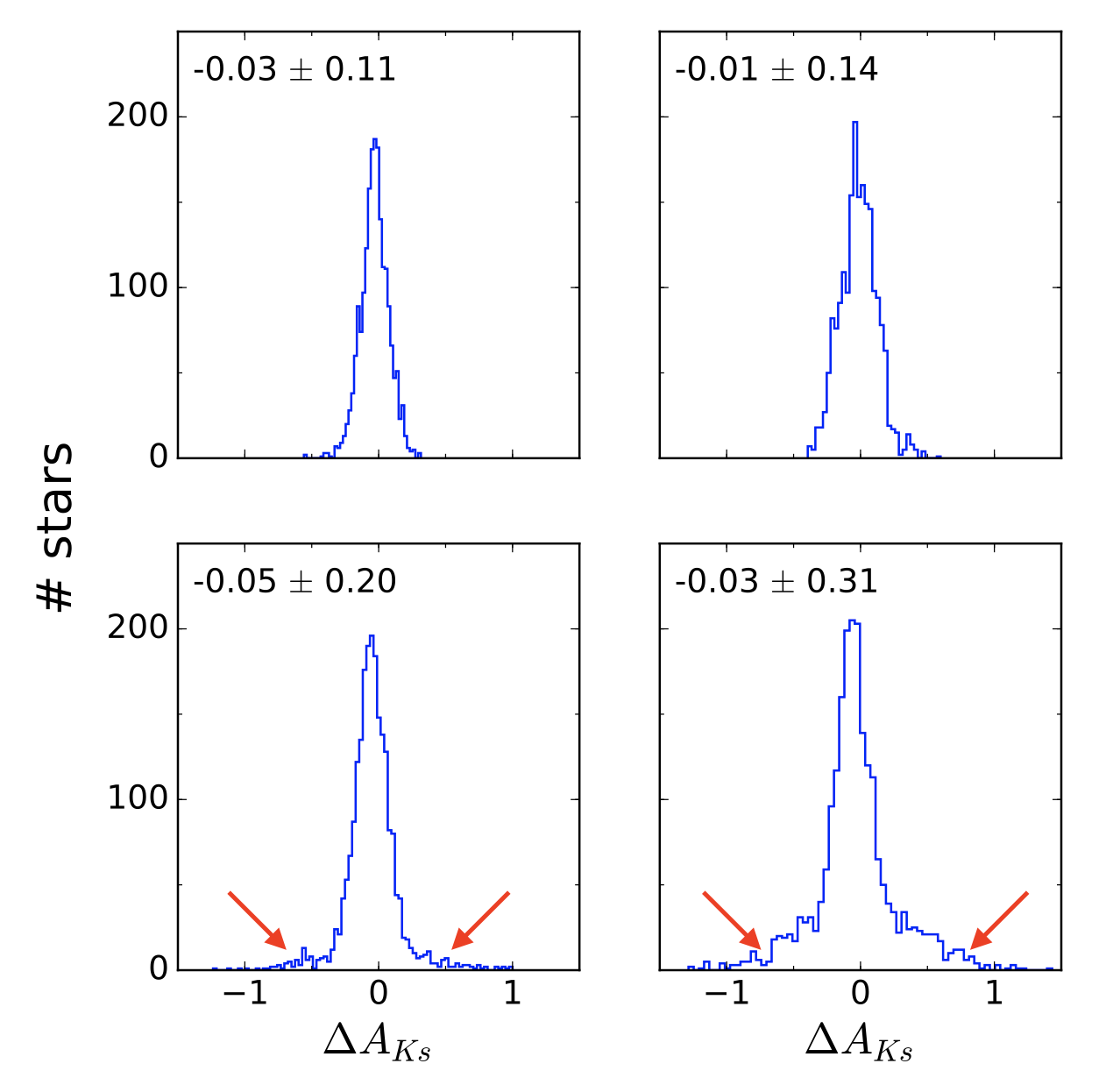}
   \caption{Comparison between initial extinction and recovered one using the extinction map (upper panels) and the star-by-star approach (lower panels). Left and right columns correspond to 15\,\% and 50\,\% of variable stars, respectively. The red arrows indicate the tails associated with variable stars for the star-by-star approach. The mean and the standard deviations of the distributions are indicated for each case.}

   \label{ext_sim_var}
    \end{figure}

To compare both approaches, we computed the difference between the real extinction and the one measured from the simulations for each star ($\Delta A_{K_s}=A_{K_s\_initial}-A_{K_s\_recovered}$). Figure\,\ref{ext_sim_var} shows the distributions for the considered variability scenarios (15\,\% and 50\,\% of variable stars). We computed the mean and the standard deviation of the distributions. We concluded that the extinction map method performs better because it suppresses uncertainties since it takes the IDW of several stars at each point and because only stars with known intrinsic colours are used to compute the map. Therefore, the extinction map method is not affected by the presence of variable stars that cause the presence of tails in the $\Delta A_{K_s}$ distributions when using the star-by-star approach. To decrease the influence of the tails and check the performance of each method, we also applied a Gaussian fit to each of the $\Delta A_{K_s}$ distributions and ended up with a larger standard deviation for the star-by-star case ($\sim 30\,\%$ and $\sim50\,\%$ larger for the case of 15\,\% and 50\,\% of variable stars, respectively). 

\begin{figure}
   \includegraphics[width=\linewidth]{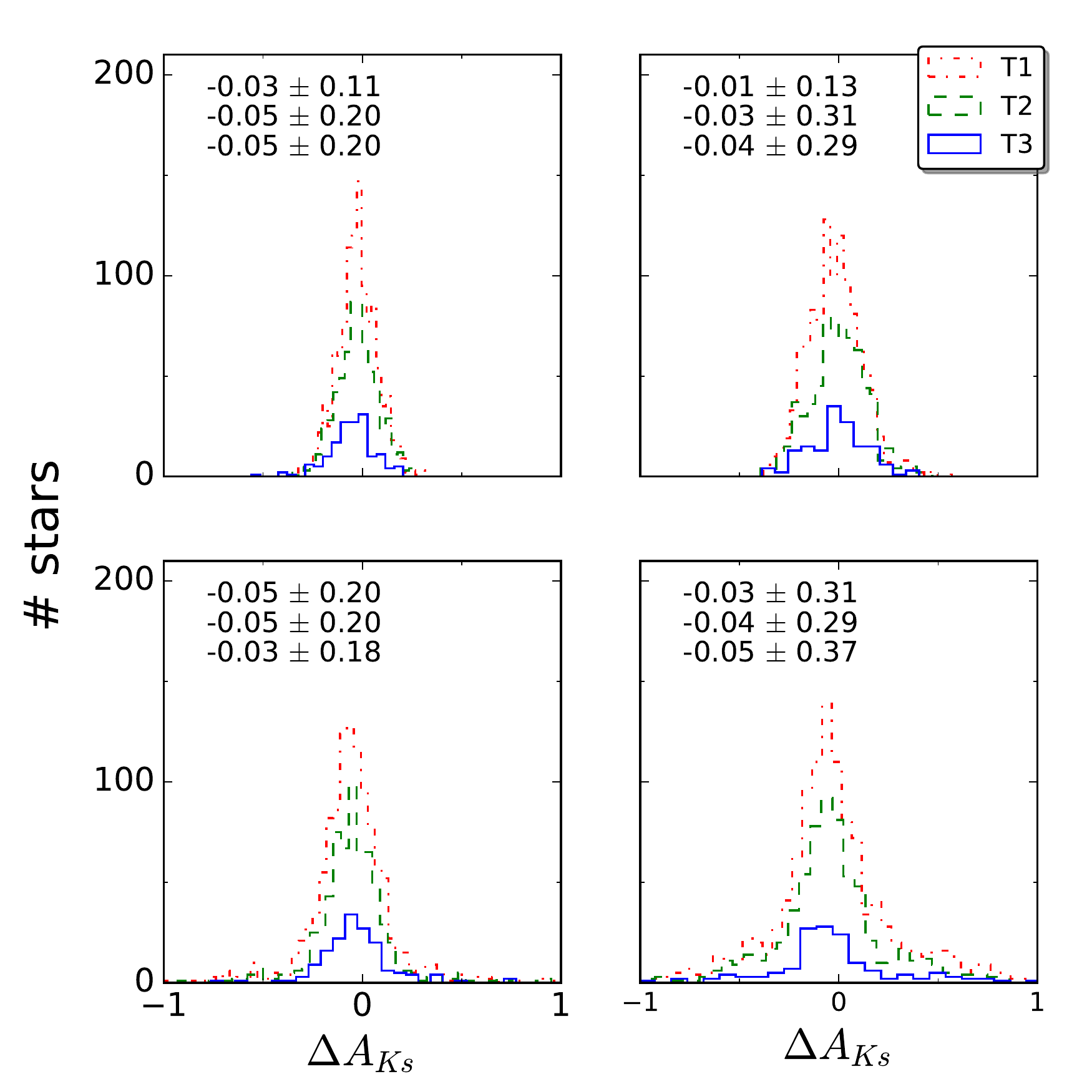}
   \caption{Stellar temperature effect on $\Delta A_{K_s}$. The upper and lower panels correspond to the extinction map and the star-by-star approach, respectively. The legend indicates the different temperature ranges considered: T1 = 2500-6500\,K, T2 = 6500-10,500\,K, and T3 = 10,500-14,500\,K. The mean and standard deviations of the distributions are included for each panel.}

   \label{ext_sim_var_T}
    \end{figure}

Moreover, we observed that the extinction-map approach shows lower mean differences between the initial values and the recovered ones. This is due to the lack of information on the spectral type of the star when applying the star-by-star de-reddening. In this sense, we also analysed the influence of the stellar temperature on the $\Delta A_{K_s}$ distributions. Figure\,\ref{ext_sim_var_T} shows that the extinction map approach works significantly better for stars with temperatures between 2500 and 6500\,K ($\sim60$\,\% of the total number of stars), whereas the star-by-star approach does not show any significant difference when comparing different stellar temperatures and is, in general, inferior to the extinction-map approach.

We concluded that the extinction map approach seems to be more appropriate, in particular when dealing with a significant fraction of variable stars. Our simple models only consider variability within an amplitude range like the one shown in Fig. 9 in \citet{2017MNRAS.470.3427D}. The inclusion of stars with larger variations would reinforce the usefulness of the extinction map approach to deal with these kind of sources.

\section{De-reddened CMDs}
\label{dereddenedCMDs}

       \begin{figure*}[h!]
   \includegraphics[scale = 0.7]{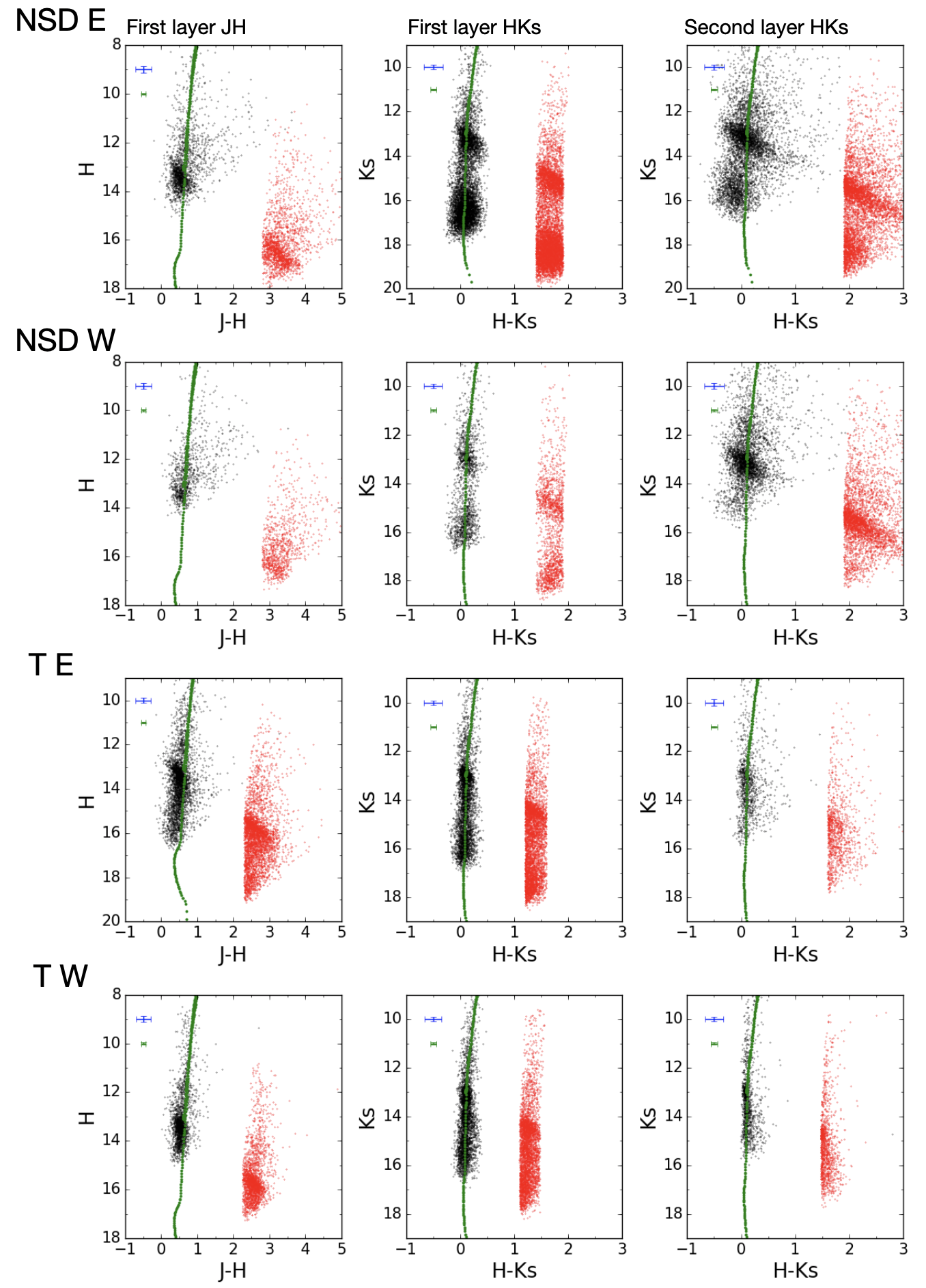}
   \caption{Colour-magnitude diagrams before and after the application of the extinction maps for the NSD E, NSD W, T E, and TW. The red and black dots are the original and the de-reddened CMDs, respectively. Only a fraction of stars are plotted given the high number of sources. The labels above each panel indicate the corresponding extinction layer. The green line depicts a Parsec isochrone of $\sim 10$ Gyr with twice solar metallicity. The blue and green error bars indicate the systematic uncertainties of the de-reddening and the ZP, respectively.}

   \label{dereddened1}
    \end{figure*}

       \begin{figure}
   \includegraphics[width=\linewidth]{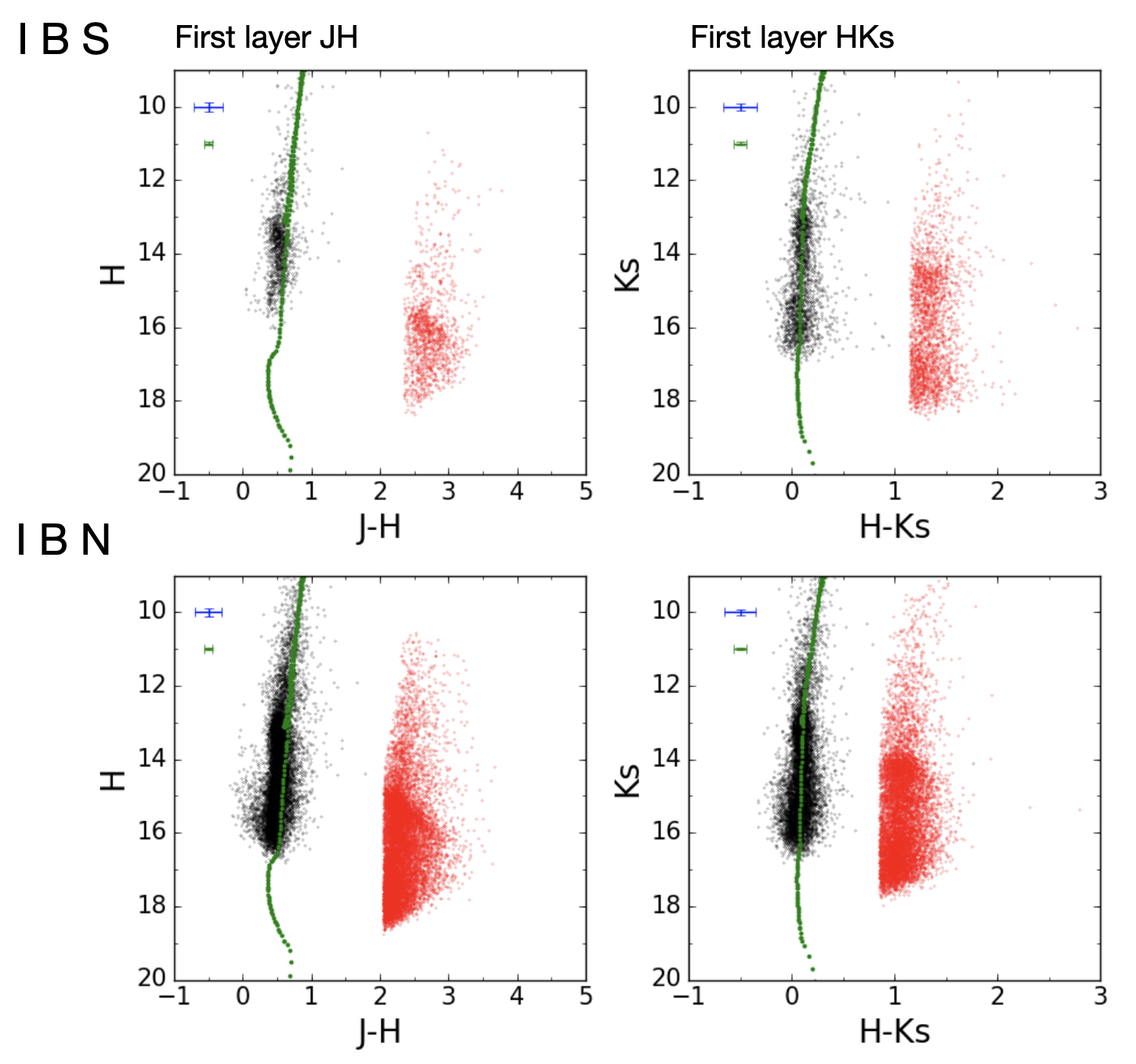}
   \caption{Colour-magnitude diagrams before and after the application of the extinction maps for the I B S and I B N. The red and black dots are the original and the de-reddened CMDs, respectively. Only a fraction of stars are plotted given the high number of sources. The labels above each panel indicate the corresponding extinction layer. The green line depicts a Parsec isochrone of $\sim 10$ Gyr with twice solar metallicity. The blue and green error bars indicate the systematic uncertainties of the de-reddening and the ZP, respectively.}

   \label{dereddened2}
    \end{figure}

\end{document}